\documentclass[aps,prl,10pt,final,twocolumn,showpacs]{revtex4}
\usepackage{graphicx}
\usepackage{array}
\usepackage{amsmath}
\usepackage{amssymb}
\usepackage{units}
\usepackage{upgreek}
\usepackage[normalem]{ulem}
\usepackage{xstring}
\let\oref\ref
\renewcommand{\ref}[1]{\protect\IfSubStr{#1}{eq:}{(\oref{#1})}{\oref{#1}}}

\newcommand{\at}[1]{\big|_{#1}}

\newcommand{\htrans}{h_{tr}}
\newcommand{\ttrans}{t_{tr}}
\newcommand{\tsc}{t_{sc}}
\newcommand{\ShN}{{S\!h_{tr}}}
\newcommand{\hmax}{h_{\scriptscriptstyle\text{peak}}}
\newcommand{\tmax}{t_{\scriptscriptstyle\text{peak}}}
\newcommand{\Dcmax}{\Delta c_{\scriptscriptstyle\text{peak}}}

\newlength{\colwidth}
\begin{document}

\title{Vertical Composition Profile During Hydrodynamic-Evaporative Film Thinning:\\
The Physics of Spin Casting Dilute Solutions}
\date{\today}

\author{Stefan Karpitschka}
\email{stefan.karpitschka@mpikg.mpg.de}
\author{Constans M. Weber}
\author{Hans Riegler}
\affiliation{Max-Planck-Institut f\"ur Kolloid- und Grenzfl\"achenforschung, Potsdam-Golm, Germany}

\begin{abstract}
We analyze the evolution of the vertical composition profile during hydrodynamic-evaporative film thinning as it typically occurs during spin casting mixtures of non-volatile solutes and volatile solvents. 
We assume that the solvent dominates the hydrodynamic-evaporative film thinning. 
The internal spatio-temporal evolution of the composition is analyzed with a diffusive-advective approach.
The analysis provides transparent physical insights into the influence of the experimental conditions on the evolution of the internal composition.
We present power laws that link the process control parameters to the composition evolution, process duration, and final solute coverage. 
The analysis reveals a characteristic Sherwood Number as fundamental process parameter.
It identifies for which stages of the process our analysis is quantitatively relevant and discloses the dominance of either diffusion or evaporation.
The analysis is valid for dilute solutions e.g., for the deposition of solute (sub)monolayers.
But it is also relevant for the deposition of thicker (polymer) films.

\end{abstract}


\maketitle

\begin{figure}
	\centering\includegraphics[width=.9\colwidth]{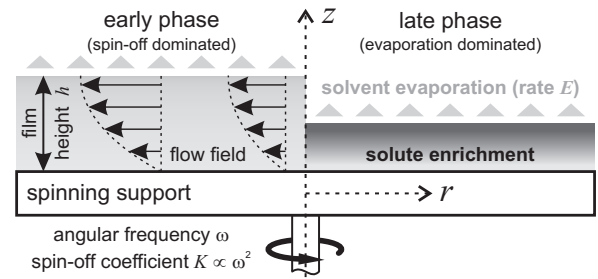}
	\caption{\label{fig:sketch}Schematics of spin-casting with processes dominating early and late stages.}
\end{figure}

\emph{Introduction.---}%
Excess amounts of liquid deposited on a spinning, planar, wettable substrate form a thinning film of uniform thickness $h$ as the liquid flows outward~\cite{Emslie:JApplPhys58}.
Evaporation of volatile film components adds to the film thinning (Fig.~\ref{fig:sketch}).
Nonvolatile components continuously enrich, eventually exceed saturation, and precipitate.
The general, fundamental physics questions in this hydraulic-evaporative (spin cast) process are:
How thins the film with time? How evolves the film composition during thinning? How much nonvolatile solute is finally deposited?

Many studies have analyzed this process experimentally%
~\cite{%
Givens:JElChemSoc79,
Daughton:JElChemSoc82,
White:JElChemSoc83,
Chen:PolEngSci83,
Weill:PolEngSci88,
McConnell:JApplPhys88,
Strong:AICHEJ89,
Skrobis:PolEngSci90,
Spangler:PolEngSci90,
Bornside:JElChemSoc91,
Peurrung:JElChemSoc91,
Britten:JApplPhys92,
Hershcovitz:MicroElRelab93,
Horowitz:JPhysIII93,
Peurrung:IEEE93,
Extrand:PolEngSci94,
Oztekin:JApplPhys95,
Vanhardeveld:ApplSurfSci95,
Birnie:OptEng95,
Gu:JApplPolSci95,
Gu:PolEngSci96,
Birnie:PhysFluids97,
Birni:SNonCrystSol97,
Horowitz:JSolGelSciTech98,
Gupta:IndEngChemRes98,
Hall:PolEngSci98,
Haas:ProcSocPhotoOptInstEng00,
Birnie:JMatRes01,
Strawhecker:Macromol01,
Walsh:ThinsolidFilms03,
Burns:ChemEngSci03,
Kim:Macromol04,
Jukes:Macromol05,
Chang:ThinSolidFilms05,
Birnie:JMatSciElectr05,
Cheung:ApplPhysLett05,
Yimsiri:ChemEngSci06,
Mohajerani:JOptoAdvMat07,
Parmar:PhysRevE09,
Parmar:PhysFluids09,
Birnie:OptLaserEng10,
Mokaraian:EPJE10,
Holloway:RheolActa10,
Toolan:JMatChemC13}
and theoretically%
~\cite{%
Emslie:JApplPhys58,
Acrivos:JApplPhys60,
Washo:IBMJResDev77,
Meyerhofer:JApplPhys78,
Matsumoto:IndEngChemFundam82,
Jenekhe:PolymEngSci83,
Jenekhe:IndEngChemFundam84,
Flack:JApplPhys84,
Sukanek:JImagTech85,
Jenekhe:ChemEngComm85,
Higgins:PhysFluids86,
Kaplon:ChemEngSci86,
Bornside:JImagTech87,
Shimoji:JapanJApplPhys87,
Yanagisawa:JApplPhys87,
Tu:JCollInterfSci87,
Stillwagon:JElChemSoc87,
Middleman:JApplPhys87,
Rehg:PhysFluids88,
Papanstasiou:JRehol88,
Lawrence:PhysFluids88,
Bornside:JApplPhys89,
Ohara:PhysFluidsA89,
Shimoji:JApplPhys89,
Matsumoto:JSMEIntJ89,
Hwang:JApplPhys89,
Ma:JApplPhys89,
Dandapat:IntJNonlinMech90,
Ma:JApplPhys90,
Stillwagon:PhysFluidsA90,
Hwang:MechResComm90,
Lawrence:PhysFluidsA90,
Tu:JCollInterfSci90,
Lawrence:JNonNewtonianFluidMech91,
Reisfeld:JApplPhys91:1,
Reisfeld:JApplPhys91:2,
Sukanek:JElChemSoc91,
Wang:IMAJApplMath91,
Kim:JApplPhys91,
Potanin:ChemEngSci92,
Stillwagon:PhysFluidsA92,
Yonkoski:JApplPhys92,
Rehg:AIChEJ92,
Bornside:JApplPhys93,
Levinson:PolEngSci93,
Dandapat:IntJNonlinMech93,
Kim:JApplPhys93,
Forcada:JCollInterfSci93,
Ma:ProbabEngEngMech94,
Spaid:JNonnewtonFluidMech94,
Borkar:PhysFluids94,
Dandapat:JPhysD94,
Gu:JElectrochemSoc95,
deSouza:CompMatSci95,
Wang:ChemEngSci95,
Yen:JChinIChE95,
Hardeveld:AppsSurfSci95,
Burgess:PhysFluids96,
Tsamopoulos:RheolActa96,
Okuzono:PhysRevLett97,
Sukanek:JElChemSoc97,
Momoniat:IntJNonlinMech98,
Dandapat:ZAngewMathMech98,
Dandapat:JPhysD99,
McKinley:PhysFluids99,
Wu:JElChemSoc99,
Wilson:JFluidMech00,
Kitamura:PhysFluids00,
Chou:JElChemSoc00,
Dandapat:PhysFluids01,
Usha:IntJNonlinMech01,
Usha:ActaMech01,
Myres:IntJNonlinMech01,
Usha:ZAngewMathPhys01,
Kitamura:PhysFluids01,
Usha:ZAngewMathMech02,
Kitamura:FluidDynRes02,
deGennes:EPJE02,
Haas:JMatSci02,
Kim:JVacSciTechnolB02,
Schubert:MatResInnov03,
Sisoev:JChemTechBio03,
Sisoev:JFluidMech03,
Dandapat:JApplPhys03,
Usha:IntJNonlinMech04,
Schwartz:PhysFluids04,
Matar:PhysFluids04,
Dandapat:PhysFluids05,
Usha:ActaMech05,
Sisoev:ChemEngSci05,
Momoniat:IntJNonlinMech05,
Wu:PhysRevE05,
Usha:FluidDynRes05,
Usha:PhysFluids05,
Myres:ChemEngProc06,
Us:PhysFluids06,
Matar:CanJChemEng06,
Tabuteau:RheolActa07,
Charpin:PhysRevE07,
Wu:SensActuatorA07,
Holloway:PhysRevE07,
Cregan:JCollInterfSci07,
Zhao:PhysFluids08,
Matar:ChemEngSci08,
Chen:MathProbEng09,
Mukhopadhyay:JPhysCondMat09,
Chen:MathProblEng10:2,
Chen:MathProblEng10:1,
Jung:IntJHMT10,
McIntyre:JFluidMech10,
Temple-Boyer:MicroEng10,
Muench:PhysFluids11,
Modhien:ApplMathMod11,
Chen:JMech11,
Dandapat:IntJNonlinMech11,
Dandapat:CommNonlinSci12,
Lin:ApplMathMod12}.
Most of them focus on the radial (in)stability of the film%
~\cite{%
Acrivos:JApplPhys60,
Givens:JElChemSoc79,
Daughton:JElChemSoc82,
White:JElChemSoc83,
Chen:PolEngSci83,
Flack:JApplPhys84,
Jenekhe:IndEngChemFundam84,
Jenekhe:ChemEngComm85,
Higgins:PhysFluids86,
Stillwagon:JElChemSoc87,
Tu:JCollInterfSci87,
McConnell:JApplPhys88,
Ma:JApplPhys89,
Hwang:JApplPhys89,
Ma:JApplPhys90,
Spangler:PolEngSci90,
Stillwagon:PhysFluidsA90,
Hwang:MechResComm90,
Kim:JApplPhys91,
Peurrung:JElChemSoc91,
Reisfeld:JApplPhys91:1,
Reisfeld:JApplPhys91:2,
Lawrence:JNonNewtonianFluidMech91,
Britten:JApplPhys92,
Potanin:ChemEngSci92,
Stillwagon:PhysFluidsA92,
Forcada:JCollInterfSci93,
Peurrung:IEEE93,
Kim:JApplPhys93,
Spaid:JNonnewtonFluidMech94,
Ma:ProbabEngEngMech94,
Oztekin:JApplPhys95,
Wang:ChemEngSci95,
Gu:JElectrochemSoc95,
Yen:JChinIChE95,
Birnie:OptEng95,
Tsamopoulos:RheolActa96,
Gupta:IndEngChemRes98,
Momoniat:IntJNonlinMech98,
McKinley:PhysFluids99,
Wu:JElChemSoc99,
Chou:JElChemSoc00,
Kitamura:PhysFluids00,
Wilson:JFluidMech00,
Kitamura:PhysFluids01,
Myres:IntJNonlinMech01,
Birnie:JMatRes01,
Strawhecker:Macromol01,
Dandapat:PhysFluids01,
Usha:IntJNonlinMech01,
Usha:ActaMech01,
Kitamura:FluidDynRes02,
Usha:ZAngewMathMech02,
Kim:JVacSciTechnolB02,
Burns:ChemEngSci03,
Sisoev:JFluidMech03,
Sisoev:JChemTechBio03,
Dandapat:JApplPhys03,
Matar:PhysFluids04,
Schwartz:PhysFluids04,
Usha:IntJNonlinMech04,
Momoniat:IntJNonlinMech05,
Dandapat:PhysFluids05,
Wu:PhysRevE05,
Usha:ActaMech05,
Sisoev:ChemEngSci05,
Matar:CanJChemEng06,
Myres:ChemEngProc06,
Us:PhysFluids06,
Charpin:PhysRevE07,
Holloway:PhysRevE07,
Tabuteau:RheolActa07,
Wu:SensActuatorA07,
Matar:ChemEngSci08,
Zhao:PhysFluids08,
Parmar:PhysFluids09,
Chen:MathProbEng09,
Parmar:PhysRevE09,
Mukhopadhyay:JPhysCondMat09,
Chen:MathProblEng10:2,
Holloway:RheolActa10,
Birnie:OptLaserEng10,
McIntyre:JFluidMech10,
Jung:IntJHMT10,
Mokaraian:EPJE10,
Modhien:ApplMathMod11,
Chen:JMech11,
Lin:ApplMathMod12},
or systems with complicated rheologies such as polymer solutions%
~\cite{%
Givens:JElChemSoc79,
Daughton:JElChemSoc82,
Jenekhe:PolymEngSci83,
White:JElChemSoc83,
Flack:JApplPhys84,
Jenekhe:IndEngChemFundam84,
Shimoji:JapanJApplPhys87,
McConnell:JApplPhys88,
Lawrence:PhysFluids88,
Weill:PolEngSci88,
Skrobis:PolEngSci90,
Lawrence:PhysFluidsA90,
Spangler:PolEngSci90,
Bornside:JElChemSoc91,
Sukanek:JElChemSoc91,
Britten:JApplPhys92,
Levinson:PolEngSci93,
Extrand:PolEngSci94,
Gu:JApplPolSci95,
deSouza:CompMatSci95,
Sukanek:JElChemSoc97,
Gupta:IndEngChemRes98,
Hall:PolEngSci98,
Haas:ProcSocPhotoOptInstEng00,
Strawhecker:Macromol01,
deGennes:EPJE02,
Walsh:ThinsolidFilms03,
Schubert:MatResInnov03,
Kim:Macromol04,
Chang:ThinSolidFilms05,
Cheung:ApplPhysLett05,
Jukes:Macromol05,
Yimsiri:ChemEngSci06,
Mohajerani:JOptoAdvMat07,
Mokaraian:EPJE10,
Temple-Boyer:MicroEng10,
Muench:PhysFluids11,
Toolan:JMatChemC13}
(the main application of spin casting is depositing polymer films) or other non-newtonian liquids%
~\cite{
Acrivos:JApplPhys60,
Matsumoto:IndEngChemFundam82,
Jenekhe:IndEngChemFundam84,
Jenekhe:ChemEngComm85,
Lawrence:JNonNewtonianFluidMech91,
Spaid:JNonnewtonFluidMech94,
Borkar:PhysFluids94,
Burgess:PhysFluids96,
Tsamopoulos:RheolActa96,
Charpin:PhysRevE07,
Tabuteau:RheolActa07,
Parmar:PhysRevE09,
Holloway:RheolActa10,
Lin:ApplMathMod12}%
.
Rather few studies%
~\cite{%
Meyerhofer:JApplPhys78,
Bornside:JImagTech87,
Lawrence:PhysFluids88,
Ohara:PhysFluidsA89,
Bornside:JApplPhys89,
Shimoji:JApplPhys89,
Lawrence:PhysFluidsA90,
Lawrence:JNonNewtonianFluidMech91,
Reisfeld:JApplPhys91:1,
Reisfeld:JApplPhys91:2,
Yonkoski:JApplPhys92,
Cregan:JCollInterfSci07,
Temple-Boyer:MicroEng10,
Muench:PhysFluids11}
aim at elucidating the general, fundamental physics of this process.
Due to different approximations/approaches there is no agreement on the final amount of solute deposition%
~\cite{%
Lawrence:PhysFluids88,
Yonkoski:JApplPhys92,
Norrman:AnRepProcChemC05,
Muench:PhysFluids11}.
Very few studies explicitly investigate the evolution of the internal film composition.
Often they assume specific cases and/or approximations such as solute "boundary layers"~\cite{Lawrence:PhysFluids88, Lawrence:PhysFluidsA90, Lawrence:JNonNewtonianFluidMech91}, or "split mechanism models"~\cite{Yonkoski:JApplPhys92}.
Some analysis is solely based on numerics~\cite{Ohara:PhysFluidsA89,Bornside:JApplPhys89, Shimoji:JApplPhys89} or combining numerics with analytical considerations assuming complicated nonlinear viscosity and diffusivity behavior~\cite{Muench:PhysFluids11}.
In any case, all those studies use vertical "initial" Peclet numbers based on inappropriate~(see \cite{Yonkoski:JApplPhys92}) "initial" film heights
~\footnote{except for~\cite{Cregan:JCollInterfSci07,Yonkoski:JApplPhys92}, which reveal very little about the compositional evolution.}.
Thus their validity ranges are ill-defined and the insight into the general physics of this interesting hydraulic-evaporative process is limited. 

Increasingly spin casting is used for fundamental nucleation and growth studies~\cite{Riegler:NatPhys07,Berg:PhysRevLett10} and for the deposition of structured (sub)monolayers (particle arrays, ``evaporation-induced self-assembly''~\cite{Brookshier:Langmuir99,Rabani:Nature03,Bigioni:NatMat06,Hanrath:Nano09,Heitsch:JPhysChem10,Klecha:JPhysChemLett10,Berg:PhysRevLett10,Johnston-Peck:Langmuir11,Marin:PhysRevLett11}).
This means dilute solutions, which behave rather ideally during most of the spin cast process.
This is the motivation and basis of our analysis.

In this report we focus on the evolution of the vertical composition profile during the hydrodynamic-evaporative film thinning.
To reveal its general aspects we neglect any complicated, non-linear solution behavior.
Thus, for the first time palpable, universal relations between the process parameters and the compositional evolution are presented.
The key process parameter, its characteristic Sherwood number $\ShN$ is introduced.
It is demonstrated that the analysis is indeed quantitatively valid for low solute concentrations but also yields valuable insights into the behavior of higher concentrated solutions including polymeric solutes.

\emph{I. Hydrodynamic-evaporative film thinning --- }%
The thinning of a Newtonian, volatile liquid film of thickness $h$ on a rotating support is described by~\cite{Meyerhofer:JApplPhys78}:
\begin{equation}
	\label{eq:dthevap}
	{dh/dt} = - 2 K\, h^3 - E\,\text{,} 
\end{equation}
with spin-off coefficient $K=\omega^2/(3\nu)$, $\omega$ = rotational speed, $\nu$ = kinematic viscosity and $E$ = evaporation rate.
The fundamental form of Eq.~\ref{eq:dthevap},
\begin{equation}
	\label{eq:dthevapfundamental}
	d\xi/d\tau = -\xi^3 - 1\,\text{,} 
\end{equation}
is obtained by rescaling $\xi=h/\htrans$ and $\tau=t/\tsc^*$ i.e., by the system inherent ``natural'' scales (see~\cite{Cregan:JCollInterfSci07}):
\begin{align}
	\label{eq:natscales}
	\htrans &= (E/2K)^{\nicefrac{1}{3}}\text{,}\\
	\tsc^* &= (2E^2 K)^{-\nicefrac{1}{3}}\text{.}
\end{align}
$\htrans$ is the  ``transition height'' where evaporative and hydrodynamic thinning are equal. 
$\tsc^*$ is the "reduced process duration" (Eq.~\ref{eq:tsc}). 

The inverse of Eq.~\ref{eq:dthevapfundamental} can be integrated%
~\cite{Cregan:JCollInterfSci07}%
~\footnote{The expression Eq. (43) presented in \cite{Reisfeld:JApplPhys91:1} is not correct!}%
:
{\small\begin{equation}
	\label{eq:th}
	\tau(\xi) = \frac{\sqrt{3}}{6}\Bigg\{   \pi
								 			+ 2\arctan\frac{1 - 2 \xi}{\sqrt{3}}
								 			+ \frac{1}{\sqrt{3}}\log \frac{1-\xi+\xi^2}{\left(1 + \xi\right)^2}
				 			\Bigg\}\text{,}
\end{equation}}%
with integration constants so that $\tau=0$ for $\xi\rightarrow\infty$. 

The total spin cast time (from $h\rightarrow\infty$ to $h=0$) is
\begin{equation}
	\label{eq:tsc}
	\tsc = (2\pi/3^{3/2})  (2E^2 K)^{-\nicefrac{1}{3}} = (2\pi/3^{3/2}) \cdot \tsc^* \text{.}
\end{equation}

\begin{figure}
	\centering\includegraphics[width=.9\colwidth]{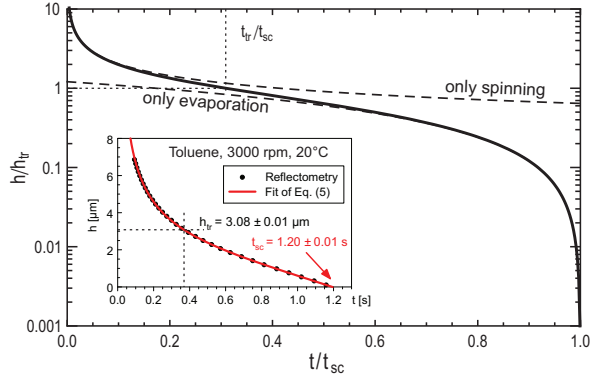}
	\caption{\label{fig:thinning}Universal thinning curve $h/\htrans=f(t/\tsc)$ (dashed lines: thinning due to spinning/evaporation only). Inset: Measured thinning curve in physical units for toluene and its fit according to the theory (Eq.~\ref{eq:th}).}
\end{figure}

Fig.~\ref{fig:thinning} shows the universal thinning curve (the inverse of Eq.~\ref{eq:th}) on scaled axes.
The transition time, $\ttrans$, at which $\htrans$ is reached, is \emph{universal} for all spin cast processes described by Eq.~\ref{eq:dthevap}:
\begin{equation}
	\label{eq:ttrans}
	\ttrans = \left( \left( 2\pi - \sqrt{3}\log 4 \right)/(4\pi)\right)\cdot \tsc\approx 0.309\cdot \tsc\text{.}
\end{equation}
Hydrodynamic film thinning ($E=0$, dashed) dominates $\approx$30\% of the spin cast time; $\approx$70\% is evaporation-dominated ($K=0$, dashed).
All this agrees well with experimental results (see inset in Fig.~\ref{fig:thinning} and supplement%
~\footnote{See supplemental material at {\tt http://link.aps.org/ supplemental/...} for experimental thinning curves, a step-by-step derivation of Eq.~\ref{eq:final}, and details on the numerical methods.}%
).

\emph{II. Solute concentration evolution --- }%
During spin casting, non-volatile solute enriches at the free surface (where the solvent evaporates) and migrates into the film via diffusion.
The spatio-temporal evolution of the solute concentration $c$ is described by
\begin{equation}
	\label{eq:conc}
	\partial_t c = D\,\partial_z^2 c + K\,z^2\,\left(3h-z\right)\,\partial_z c\text{,}
\end{equation}
with boundary conditions%
\begin{equation}
	D\,\partial_z c\at{z=h} = E\, c\at{z=h}\text{,}
	\quad\quad\quad
	\partial_z c\at{z=0} = 0\text{.}
\end{equation}

The advective term in Eq.~\ref{eq:conc} is derived from the radial velocity field $u(r,z)$ of a Newtonian fluid rotating with its solid support (no slip), with a free surface at the top (no stress)~\cite{Emslie:JApplPhys58}:
\begin{equation}
	u(r,z) = 3K\, r\, z\,\left( h - {z/2} \right)\text{.}
\end{equation}
$r$ and $z$ are the radial respectively vertical coordinates. 
The radial volumetric flux $\phi\, dz$ is
\begin{align}
	\phi\,dz &= 2\pi\, r\, u(r,z)\, dz = 6\pi\,K\, r^2 z\left( h - z/2 \right)dz\text{.}
\end{align}
With the continuity equation this yields the thinning-induced vertical motion of the horizontal stream lines:
\begin{equation}
	\label{eq:dZdt}
	dZ/dt = -1/(2\pi\, r)\,\int_0^z \partial_r \phi\, dz' = - K\,z^2\,\left(3h-z\right)\text{.}
\end{equation}

In order to solve Eq.~\ref{eq:conc}, $h(t)$ is required but not known explicitly, only its inverse, $t(h)$ (Eq.~\ref{eq:th}). 
Since $h(t)$ respectively $\tau(\xi)$ are bijective, the time variable can be changed to $\xi$:
$\partial_t c = d\tau/dt\cdot d\xi/d\tau\cdot \partial_{\xi} c$ \hbox{$= - (2E^2 K)^{\nicefrac{1}{3}}( \xi^3 + 1 )\,\partial_{\xi} c$}. 
Rescaling $y=z/h$ ($y\in [0, 1]$, from substrate to surface) avoids the moving boundary.
This leads to:
\begin{equation}
	\label{eq:final}
	\partial_{\xi} c = - \frac{\partial_y^2 c}{\ShN\,\xi^2(\xi^3 + 1)}
								     - \left\{\frac{(\xi\, y)^2(3 - y)}{2(\xi^3 + 1)} - \frac{y}{\xi}\right\}\partial_y c\text{,}
\end{equation}
\begin{equation}
	\partial_y c\at{y=1} = \ShN\,\xi\,c\at{y=1}\text{,}
	\quad\quad\quad
	\partial_y c\at{y=0} = 0\text{,}
\end{equation}
where the Sherwood number, $\ShN$, parameterizes the ratio of evaporative to diffusive mass transport on the characteristic length scale of the system, $\htrans$:
\begin{equation}
	\label{eq:Sh}
	\ShN=E\,\htrans/D=E^{\nicefrac{4}{3}}\,(2K)^{-\nicefrac{1}{3}}/D\text{.}
\end{equation}
By scaling $c$ with the initial solute concentration $c_0$ the initial condition is $c\at{\xi\rightarrow\infty} = 1$. 
Thus the system is parametrized completely by $\ShN$.
Eq.~\ref{eq:final} can be solved numerically with $\xi$ as independent variable.
Eq.~\ref{eq:th} then provides $\tau$ as function of $\xi$ i.e., finally, $c(t,z)$.

\begin{figure}
	\centering\includegraphics[width=.87\colwidth]{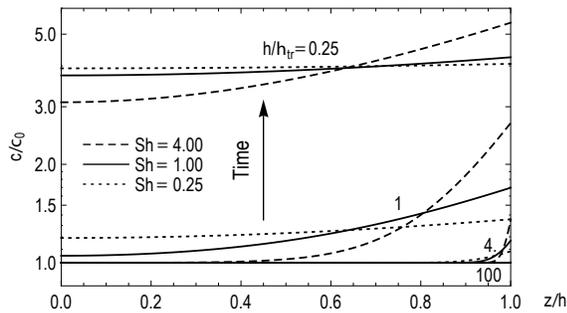}
	\caption{\label{fig:ResP}Solute concentration profiles (resulting from Eq.~\ref{eq:final}) for three different Sherwood numbers $\ShN$ at three different moments respectively heights $h/\htrans$.}
\end{figure}

\emph{III. General aspects of the concentration evolution --- }%
Based on the solution of Eq.~\ref{eq:final} the impact of $\ShN$ (i.e., of the individual system parameters $K$, $E$, and $D$) is now analyzed.
Fig.~\ref{fig:ResP} shows profiles of $c$ during film thinning for $\ShN$ larger and smaller than $1$ (i.e., convection dominating over diffusion and vice versa) and for film heights larger and smaller than $\htrans$, respectively. 
It reveals the competition between evaporative enrichment, spin-off, and diffusive dilution. 
In general, larger $\ShN$ means more pronounced gradients in $c$. 
If $h\gg\htrans$, solute enrichment occurs only locally near the free surface and $c\approx c_0$ near the substrate. 
If $h\ll\htrans$, $c$ also increases near the substrate. 
$\ShN=1$ and $h=\htrans$ mark the transition: 
The solute gradient just reaches the film/substrate interface and $c$ just begins to increase globally. 

\begin{figure}[b]
	\centering\includegraphics[width=0.9\colwidth]{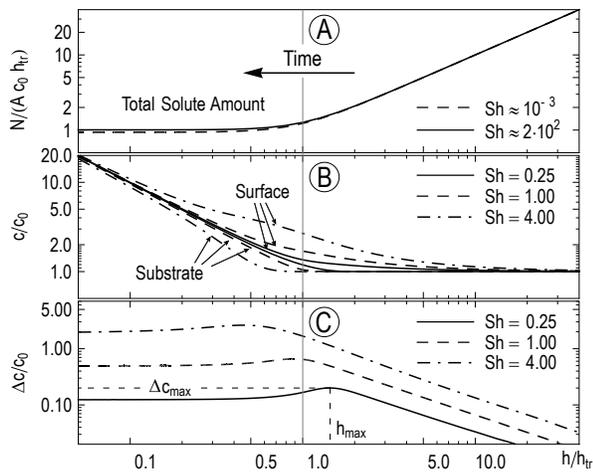}
	\caption{\label{fig:Res_N(h)+c(h)}For various Sherwood numbers $\ShN$ are shown as function of the reduced film thickness $h/\htrans$: A) Total solute amount $N$ per per unit area $A$ (scaled by $c_0\htrans$); B) Concentrations $c$ at the surface respectively film/substrate interface (scaled by $c_0$); C) $\Delta c$, the difference between the surface and the substrate/film interface concentrations, respectively (scaled by $c_0$).} 
\end{figure}

Fig.~\ref{fig:Res_N(h)+c(h)} shows as function of $h/\htrans$: 
A) The total solute amount $N$ per unit area $A$ ($N/A=\int_0^h c\, dz$); 
B) $c/c_0$ at the surface respectively substrate/film interface; 
C) The difference between the surface and the substrate/film interface concentrations, $\Delta c/c_0$ i.e., the $\emph{relative}$ enrichment.
Both, $h=\htrans$ and $\ShN = 1$ mark transitions between distinctly different behaviors. 
I) For $h\gg\htrans$ spin-off dominates film thinning. 
Therefore, globally $c/c_0$ remains approximately constant and $N/A$ decreases.
Nevertheless, there is a surficial solute enrichment due to evaporation. 
But this $\Delta c/{c_0}$ only becomes substantial for $\ShN>1$. It has a maximum at $h=\hmax$ ($\hmax\approx\htrans$ for $\ShN\approx1$).
II) For $h\ll\htrans$, evaporation dominates. 
Solvent loss leads to increasing $c/{c_0}$ while $N/A$ and $\Delta c/{c_0}$ remain constant (because spin-off becomes negligible). 
Remarkably, for $h<\htrans$, $N/(Ac_0\htrans)$ remains approximately constant and independent from $\ShN$ whereas $\Delta c/{c_0}$ increases with increasing $\ShN$. 
 
\begin{figure}[t]
	\centering\includegraphics[width=0.9\colwidth]{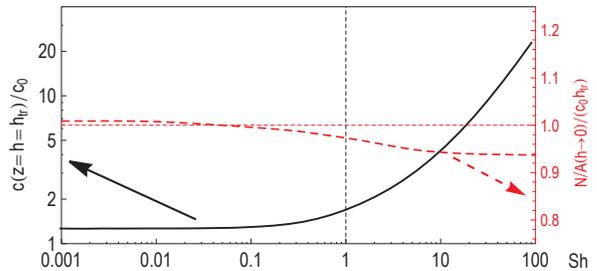}
	\caption{\label{fig:totalcovandconc}Normalized final solute coverage, $N(h\rightarrow0)/(Ac_0\htrans)$ and $c/c_0$ at the film surface  for $h=\htrans$ as function $\ShN$.}
\end{figure}

Fig.~\ref{fig:totalcovandconc} shows $c(z\!=\!h\!=\!\htrans)/c_0$, the concentration at the free surface for $h = \htrans$, and $N(h\rightarrow0)/(A c_0 \htrans)$, the rescaled final coverage. Both are plotted as function of $\ShN$. For $\ShN\ll 1$ diffusion dominates and $c$ is mostly homogeneous (Fig.~\ref{fig:ResP}). 
In this case $c_0 E\tsc$ is the total amount of solute that is \emph{not} spun off with the solvent. 
Distributing this into a film with $\htrans$ yields $c/c_0=E\tsc/\htrans=2\pi/3^{3/2}\approx 1.2$ (Fig.~\ref{fig:totalcovandconc}). 
The final solute coverage for $\ShN<1$ is (in agreement with~\cite{Bornside:JApplPhys93})
\begin{equation}
	\label{eq:Nnumerical}
	\Gamma = N(h\rightarrow0)/A \approx c_0\,\htrans\approx 0.8\, c_0\, (K/E)^{-\nicefrac{1}{3}}\text{.}
\end{equation}
For $\ShN\gg1$ it is only $\approx6\%$ lower.

\begin{figure}
	\centering\includegraphics[width=0.85\colwidth]{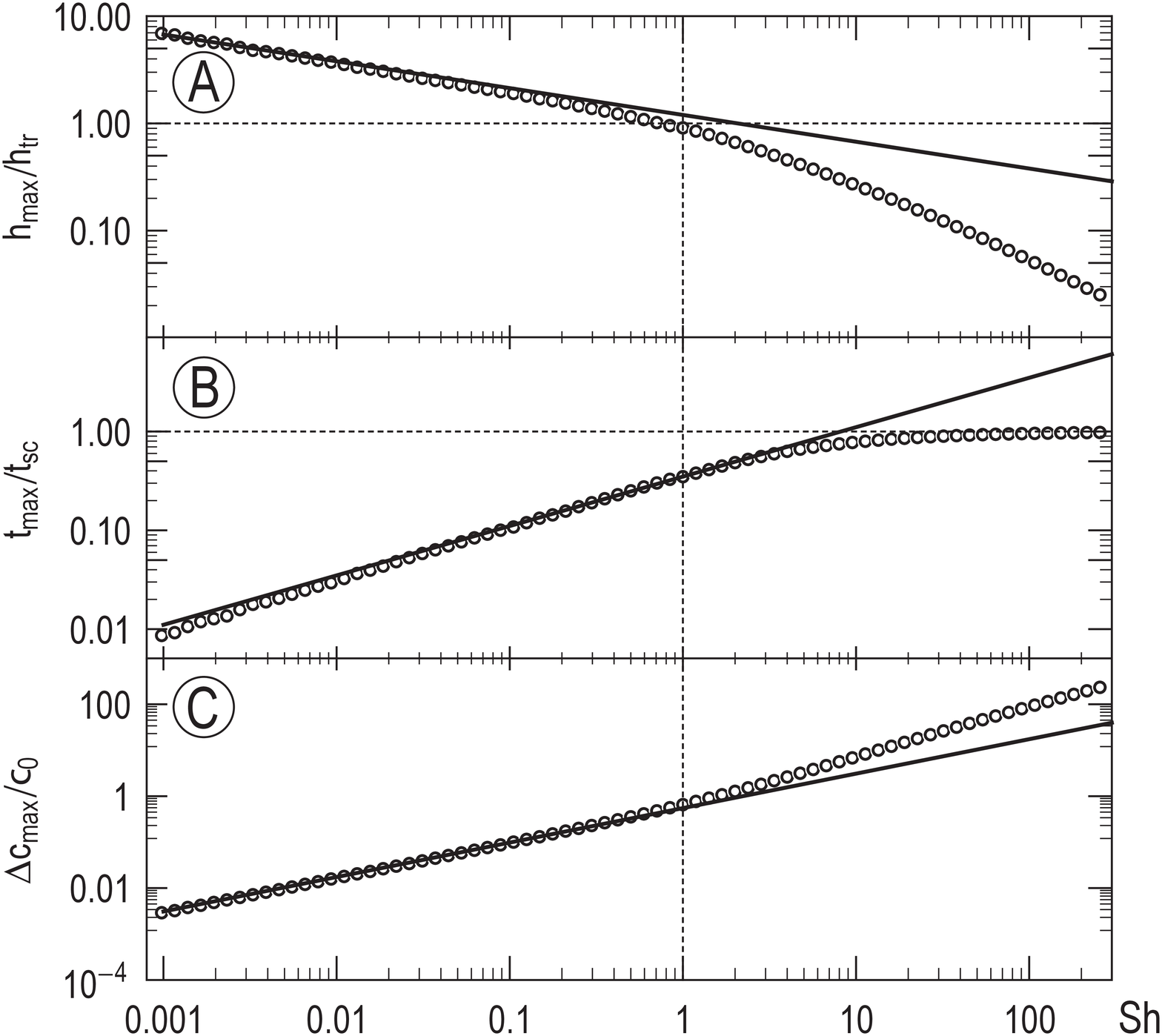}
	\caption{\label{fig:ResDCmax} $\Dcmax/c_0$ as a function of $\ShN$ and related spatio-temporal properties ($\hmax$=$h\at{\Dcmax}$ and $\tmax$=$t\at{\Dcmax}$). The symbols indicate numerical results, the lines the underlying power laws for $\ShN<1$.}
\end{figure}

\emph{IV. Relative surficial enrichment maximum --- }%
Fig.~\ref{fig:ResDCmax} shows $\Dcmax/c_0$ and its spatio-temporal properties as function of $\ShN$. 
Symbols denote the results from the numerical analysis. 
The solid lines show power laws for $\ShN<1$ which are rationalized by analyzing the underlying processes.

Panel~A) shows $\hmax$ rescaled by $\htrans$ i.e., the film thickness at which $\Delta c=\Dcmax$ as function of $\ShN$. 
$\Dcmax$ emerges from the competition between evaporative enrichment, spin-off, and diffusional equilibration. 
Evaporation and spin-off dominate at opposite ranges of $h$. 
For large $h$, surficial spin-off efficiently suppresses enrichment: 
$\Dcmax$ requires $2K\hmax^3<E$ (from Eq.~\ref{eq:dthevap}). 
However, for small $h$, diffusional equilibration is fast, so $\Dcmax$ requires $D/\hmax<E$. 
Optimizing both conditions simultaneously ($E$ is linear and protagonistic in both cases, and therefore cancels out) reveals the same power law as numerics (which supplies the prefactor):
\begin{equation}
	\label{eq:hmax}
	\hmax \approx (D/K)^{\nicefrac{1}{4}}\approx1.2\,\htrans\,\ShN^{-\nicefrac{1}{4}}\text{.}
\end{equation}

Panel~B) shows the time $\tmax$ rescaled by $\tsc$ i.e., the time at which $\Delta c=\Dcmax$. 
Before reaching $\hmax$, thinning is dominated by spin-off. 
Hence $\tmax$ can be estimated by inserting $\hmax$ into $h=(2K\,t)^{-\nicefrac{1}{2}}$ (the solution to Eq.~\ref{eq:dthevap} with $E=0$):
\begin{equation}
	\label{eq:tmax}
	\tmax/\tsc \approx 0.31\, E^{\nicefrac{2}{3}}K^{-\nicefrac{1}{6}}D^{-\nicefrac{1}{2}} \approx 0.35\,\sqrt{\ShN}\text{.}
\end{equation}

Panel~C presents $\Dcmax/c_0$, reflecting the balance between evaporative enrichment and diffusive equilibration i.e., $\ShN\at{\hmax} = E\,\hmax/D = \ShN^{\nicefrac{3}{4}}$. 
With Eq.~\ref{eq:hmax} this means:
\begin{equation}
	\label{eq:dcmax}
	\Dcmax/c_0 \approx 0.46\, E\, K^{-\nicefrac{1}{4}}D^{-\nicefrac{3}{4}}\approx 0.55\, \ShN^{\nicefrac{3}{4}} \text{.}
\end{equation}

\emph{Discussion and Conclusions --- }%
Section I introduces the system-specific fundamental length and time scales ($\htrans$, $\tsc^*$ ) for the spin casting process of an ideal Newtonian volatile liquid. 
These reduce the general spin cast equation (Eq.~\ref{eq:dthevap}) to its fundamental form~\cite{Cregan:JCollInterfSci07} and lead to a universal film thinning behavior (Eq.~$\ref{eq:th}$).
The total spin coating time, $\tsc$ is calculated (Eq.~$\ref{eq:tsc}$) as function of the process parameters ($E$, $K$, $D$).
For any combination of these parameters, in the first $30\%$ of the process time, thinning is governed by hydrodynamics.
During $70\%$ of the time evaporative thinning dominates until complete drying. 

Sections II through IV analyze the spin cast process of a mixture of a volatile solvent and a non-volatile solute assuming constant process parameters $E$, $K$, and $D$.

It is found that the spatio-temporal evolution of the solute concentration within the thinning film is universally characterized by a Sherwood number, $\ShN$, scaled to the system-inherent fundamental length, $\htrans$ (it reflects the competition between evaporative solute enrichment and diffusional dilution at $\htrans$). 
For $\ShN<1$ (diffusion dominates) the spatio-temporal occurrence of the relative surficial enrichment maximum is related to the process parameters via universal power laws  (Eqs.~\ref{eq:hmax}, \ref{eq:tmax}, and \ref{eq:dcmax}, see Fig.~\ref{fig:ResDCmax}).
These findings are rationalized semiquantitatively with the underlying physics.
At last, the final solute coverage $\Gamma$ is calculated from the process parameters (Eq.~\ref{eq:Nnumerical}).  

To examine the relevance of our analysis for real cases, where $E$, $K$, and $D$ are not necessarily always constant (e.g. depending on the solute concentration), we assume a typical solvent (e.g. toluene) with
molar mass $M_{\scriptscriptstyle\text{S}} \approx \unit[0.1]{kg/mol}$,
density $\rho_{\scriptscriptstyle\text{S}}\approx \unit[10^3]{kg/m^3}$,
$\nu=6\cdot \unit[10^{-7}]{m^{-2}s^{-1}}$,
and $E=\unit[10^{-6}]{m/s}$ [18]. 
We consider two examples: 
A) A typical polymer solution~\cite{Norrman:AnRepProcChemC05}; B) A nanoparticle solution for submonolayer deposition.

Case A): Polymer with
$M_{\scriptscriptstyle\text{P}}=\unit[100]{kg/mol}$,
$\rho_{\scriptscriptstyle\text{P}}=\unit[10^3]{kg m^{-3}}$,
and $c_{0,\scriptscriptstyle\text{P}}=\unit[10]{kg m^{-3}}$ (i.e., mole fractions $x_{0,\scriptscriptstyle\text{P}}= 10^{-6}$ and $x_{0,\scriptscriptstyle\text{S}}\approx 1$), 
$D_{\scriptscriptstyle\text{P}}=\unit[10^{-11}]{m^{2}s^{-1}}$ \cite{Rauch:PhysRevLett02}
and $K=\unit[10^{11}]{m^{-2}s^{-1}}$ ($\approx\unit[4000]{rpm}$). 
This yields: 
$\htrans\approx\unit[1.7 \cdot 10^{-6}]{m}$ (Eq.~$\ref{eq:natscales}$), 
$\ShN\approx 0.17$ (Eq.~$\ref{eq:Sh}$),
$c_{\scriptscriptstyle\text{P}}(\htrans)/c_{0,\scriptscriptstyle\text{P}}\approx1.2$ (Figs.~\ref{fig:ResP} and~\ref{fig:totalcovandconc}) and
$\Gamma_p\approx \unit[17*10^{-6}]{kg m^{-2}}$ (Eq.~$\ref{eq:Nnumerical}$) i.e., a final film thickness of $\approx \unit[17]{nm}$. This is in reasonable agreement with experimental results~\cite{Norrman:AnRepProcChemC05}. 

Case B): Spheres with 
radius $r_{\scriptscriptstyle\text{NP}}=\unit[36]{nm}$.
$D_{\scriptscriptstyle\text{NP}}=\unit[10^{-11}]{m^{2}s^{-1}}$ (estimation: Stokes-Einstein) as in case A) and thus $\htrans$ and $\ShN$ are identical to case A). 
A $50\%$ monolayer coverage means $\Gamma_{\scriptscriptstyle\text{NP}}\approx \unit[10^{14}]{m^{-2}}$ i.e., $c_{0,\scriptscriptstyle\text{NP}}\approx \unit[5\cdot10^{-5}]{mol/m^3}$ ($x_{0,\scriptscriptstyle\text{NP}}\approx5\cdot 10^{-9}$, $x_{0,\scriptscriptstyle\text{S}}\approx1$) with Eq.~$\ref{eq:Nnumerical}$.
 
In both cases the initial solute mole fraction is small.
Therefore film thinning to $\htrans$ will follow Eq.~\ref{eq:dthevap} with $K$ and $E$ remaining approximately constant because thinning is hydrodynamically dominated and the solute concentration barely changes. 
Hence $\Gamma$ can be calculated with Eq.~$\ref{eq:Nnumerical}$, which is not affected by changes in $K$ and
$E$ occurring $\emph{after}$ reaching $\htrans$.

For both examples $\ShN<1$. 
Therefore our analysis regarding $\hmax$, $\tmax$, and $\Dcmax$ is relevant even $\emph{quantitatively}$ because all three maxima occur at $h>\hmax$ i.e., the (surficial) solute concentration has not yet increased substantially above $c_{0}$ (Figs.~\ref{fig:totalcovandconc} and \ref{fig:ResDCmax}).
Accordingly, also Eqs.~\ref{eq:hmax}, \ref{eq:tmax} and \ref{eq:dcmax} are applicable.

Of course, evaporative film thinning at $h<\hmax$ will eventually increase the solute concentration and thus change $E$ and $K$.
However, in particular for "evaporation-structured" submonolayer particle array deposition, non-ideality will only become relevant for $h\ll\htrans$, when $h$ is already in the range of $r_{\scriptscriptstyle\text{NP}}$, because typically $x_{0,\scriptscriptstyle\text{NP}}\ll1$.
For low solubilities our analysis reveals quantitatively whether solute aggregation occurs first at the top surface (if evaporative solute enrichment dominates) or globally homogeneous (or heterogeneously at the substrate) if diffusive equilibration is more efficient at the corresponding film thickness.
Last not least, our analysis predicts a quantifiable behavior based on measurable parameters of the $\emph{initial}$ solution.
In particular the predicted $\Gamma$ and $h(t)$  [18] are easily measurable. 
Thus not only the validity of the approach can be evaluated.
It is possible to specifically address the influence of individual parameters of the multi-parameter spin casting process. 
For instance, parameters not related to mixing properties (e.g., temperature changes induced by evaporation) can be probed, because a sufficient reduction of $c_0$ will per definition render the solution behaving approximately ``ideally''.
All this proposes our approach as a future basis for incorporating specific (non)-linear properties of specific systems. 
Due to the predominantly analytic approach such a refined approach still will reveal general physical insights.

\acknowledgments{We thank Andreas Vetter and John Berg for scientific discussions, Helmuth M\"ohwald for scientific advice and general support. SK was funded by DFG Grant RI529/16-1, CMW by IGRTG 1524.}


\begin{thebibliography}{173}
\expandafter\ifx\csname natexlab\endcsname\relax\def\natexlab#1{#1}\fi
\expandafter\ifx\csname bibnamefont\endcsname\relax
  \def\bibnamefont#1{#1}\fi
\expandafter\ifx\csname bibfnamefont\endcsname\relax
  \def\bibfnamefont#1{#1}\fi
\expandafter\ifx\csname citenamefont\endcsname\relax
  \def\citenamefont#1{#1}\fi
\expandafter\ifx\csname url\endcsname\relax
  \def\url#1{\texttt{#1}}\fi
\expandafter\ifx\csname urlprefix\endcsname\relax\def\urlprefix{URL }\fi
\providecommand{\bibinfo}[2]{#2}
\providecommand{\eprint}[2][]{\url{#2}}

\bibitem[{\citenamefont{Emslie et~al.}(1958)\citenamefont{Emslie, Bonner, and
  Peck}}]{Emslie:JApplPhys58}
\bibinfo{author}{\bibfnamefont{A.~G.} \bibnamefont{Emslie}},
  \bibinfo{author}{\bibfnamefont{F.~T.} \bibnamefont{Bonner}},
  \bibnamefont{and} \bibinfo{author}{\bibfnamefont{L.~G.} \bibnamefont{Peck}},
  \bibinfo{journal}{J. Appl. Phys.} \textbf{\bibinfo{volume}{29}},
  \bibinfo{pages}{858} (\bibinfo{year}{1958}).

\bibitem[{\citenamefont{Givens and Daughton}({1979})}]{Givens:JElChemSoc79}
\bibinfo{author}{\bibfnamefont{F.}~\bibnamefont{Givens}} \bibnamefont{and}
  \bibinfo{author}{\bibfnamefont{W.}~\bibnamefont{Daughton}},
  \bibinfo{journal}{{J. Electrochem. Soc.}} \textbf{\bibinfo{volume}{{126}}},
  \bibinfo{pages}{269} (\bibinfo{year}{{1979}}).

\bibitem[{\citenamefont{Daughton and Givens}({1982})}]{Daughton:JElChemSoc82}
\bibinfo{author}{\bibfnamefont{W.}~\bibnamefont{Daughton}} \bibnamefont{and}
  \bibinfo{author}{\bibfnamefont{F.}~\bibnamefont{Givens}},
  \bibinfo{journal}{{J. Electrochem. Soc.}} \textbf{\bibinfo{volume}{{129}}},
  \bibinfo{pages}{173} (\bibinfo{year}{{1982}}).

\bibitem[{\citenamefont{White}({1983})}]{White:JElChemSoc83}
\bibinfo{author}{\bibfnamefont{L.}~\bibnamefont{White}}, \bibinfo{journal}{{J.
  Electrochem. Soc.}} \textbf{\bibinfo{volume}{{130}}}, \bibinfo{pages}{1543}
  (\bibinfo{year}{{1983}}).

\bibitem[{\citenamefont{Chen}({1983})}]{Chen:PolEngSci83}
\bibinfo{author}{\bibfnamefont{B.}~\bibnamefont{Chen}},
  \bibinfo{journal}{{Polym. Eng. Sci.}} \textbf{\bibinfo{volume}{{23}}},
  \bibinfo{pages}{399} (\bibinfo{year}{{1983}}).

\bibitem[{\citenamefont{Weill and Dechenaux}({1988})}]{Weill:PolEngSci88}
\bibinfo{author}{\bibfnamefont{A.}~\bibnamefont{Weill}} \bibnamefont{and}
  \bibinfo{author}{\bibfnamefont{E.}~\bibnamefont{Dechenaux}},
  \bibinfo{journal}{{Polym. Eng. Sci.}} \textbf{\bibinfo{volume}{{28}}},
  \bibinfo{pages}{945} (\bibinfo{year}{{1988}}).

\bibitem[{\citenamefont{McConnell}({1988})}]{McConnell:JApplPhys88}
\bibinfo{author}{\bibfnamefont{W.}~\bibnamefont{McConnell}},
  \bibinfo{journal}{{J. Appl. Phys.}} \textbf{\bibinfo{volume}{{64}}},
  \bibinfo{pages}{2232} (\bibinfo{year}{{1988}}).

\bibitem[{\citenamefont{Strong and Middleman}({1989})}]{Strong:AICHEJ89}
\bibinfo{author}{\bibfnamefont{L.}~\bibnamefont{Strong}} \bibnamefont{and}
  \bibinfo{author}{\bibfnamefont{S.}~\bibnamefont{Middleman}},
  \bibinfo{journal}{{AICHE J.}} \textbf{\bibinfo{volume}{{35}}},
  \bibinfo{pages}{1753} (\bibinfo{year}{{1989}}).

\bibitem[{\citenamefont{Skrobis et~al.}({1990})\citenamefont{Skrobis, Denton,
  and Skrobis}}]{Skrobis:PolEngSci90}
\bibinfo{author}{\bibfnamefont{K.}~\bibnamefont{Skrobis}},
  \bibinfo{author}{\bibfnamefont{D.}~\bibnamefont{Denton}}, \bibnamefont{and}
  \bibinfo{author}{\bibfnamefont{A.}~\bibnamefont{Skrobis}},
  \bibinfo{journal}{{Polym. Eng. Sci.}} \textbf{\bibinfo{volume}{{30}}},
  \bibinfo{pages}{193} (\bibinfo{year}{{1990}}).

\bibitem[{\citenamefont{Spangler et~al.}({1990})\citenamefont{Spangler,
  Torkelson, and Royal}}]{Spangler:PolEngSci90}
\bibinfo{author}{\bibfnamefont{L.}~\bibnamefont{Spangler}},
  \bibinfo{author}{\bibfnamefont{J.}~\bibnamefont{Torkelson}},
  \bibnamefont{and} \bibinfo{author}{\bibfnamefont{J.}~\bibnamefont{Royal}},
  \bibinfo{journal}{{Polym. Eng. Sci.}} \textbf{\bibinfo{volume}{{30}}},
  \bibinfo{pages}{644} (\bibinfo{year}{{1990}}).

\bibitem[{\citenamefont{Bornside et~al.}({1991})\citenamefont{Bornside,
  Macosko, and Scriven}}]{Bornside:JElChemSoc91}
\bibinfo{author}{\bibfnamefont{D.}~\bibnamefont{Bornside}},
  \bibinfo{author}{\bibfnamefont{C.}~\bibnamefont{Macosko}}, \bibnamefont{and}
  \bibinfo{author}{\bibfnamefont{L.}~\bibnamefont{Scriven}},
  \bibinfo{journal}{{J. Electrochem. Soc.}} \textbf{\bibinfo{volume}{{138}}},
  \bibinfo{pages}{317} (\bibinfo{year}{{1991}}).

\bibitem[{\citenamefont{Peurrung and Graves}({1991})}]{Peurrung:JElChemSoc91}
\bibinfo{author}{\bibfnamefont{L.}~\bibnamefont{Peurrung}} \bibnamefont{and}
  \bibinfo{author}{\bibfnamefont{D.}~\bibnamefont{Graves}},
  \bibinfo{journal}{{J. Electrochem. Soc.}} \textbf{\bibinfo{volume}{{138}}},
  \bibinfo{pages}{2115} (\bibinfo{year}{{1991}}).

\bibitem[{\citenamefont{Britten and Thomas}({1992})}]{Britten:JApplPhys92}
\bibinfo{author}{\bibfnamefont{J.}~\bibnamefont{Britten}} \bibnamefont{and}
  \bibinfo{author}{\bibfnamefont{I.}~\bibnamefont{Thomas}},
  \bibinfo{journal}{{J. Appl. Phys.}} \textbf{\bibinfo{volume}{{71}}},
  \bibinfo{pages}{972} (\bibinfo{year}{{1992}}).

\bibitem[{\citenamefont{Hershcovitz and
  Klein}({1993})}]{Hershcovitz:MicroElRelab93}
\bibinfo{author}{\bibfnamefont{M.}~\bibnamefont{Hershcovitz}} \bibnamefont{and}
  \bibinfo{author}{\bibfnamefont{I.}~\bibnamefont{Klein}},
  \bibinfo{journal}{{Microelectron. Reliab.}} \textbf{\bibinfo{volume}{{33}}},
  \bibinfo{pages}{869} (\bibinfo{year}{{1993}}).

\bibitem[{\citenamefont{Horowitz et~al.}({1993})\citenamefont{Horowitz,
  Yeatman, Dawnay, and Fardad}}]{Horowitz:JPhysIII93}
\bibinfo{author}{\bibfnamefont{F.}~\bibnamefont{Horowitz}},
  \bibinfo{author}{\bibfnamefont{E.}~\bibnamefont{Yeatman}},
  \bibinfo{author}{\bibfnamefont{E.}~\bibnamefont{Dawnay}}, \bibnamefont{and}
  \bibinfo{author}{\bibfnamefont{A.}~\bibnamefont{Fardad}},
  \bibinfo{journal}{{J. Phys. III}} \textbf{\bibinfo{volume}{{3}}},
  \bibinfo{pages}{2059} (\bibinfo{year}{{1993}}).

\bibitem[{\citenamefont{Peurrung and Graves}({1993})}]{Peurrung:IEEE93}
\bibinfo{author}{\bibfnamefont{L.}~\bibnamefont{Peurrung}} \bibnamefont{and}
  \bibinfo{author}{\bibfnamefont{D.}~\bibnamefont{Graves}},
  \bibinfo{journal}{{IEEE Trans. Semicond. Manuf.}}
  \textbf{\bibinfo{volume}{{6}}}, \bibinfo{pages}{72} (\bibinfo{year}{{1993}}).

\bibitem[{\citenamefont{Extrand}({1994})}]{Extrand:PolEngSci94}
\bibinfo{author}{\bibfnamefont{C.}~\bibnamefont{Extrand}},
  \bibinfo{journal}{{Polym. Eng. Sci.}} \textbf{\bibinfo{volume}{{34}}},
  \bibinfo{pages}{390} (\bibinfo{year}{{1994}}).

\bibitem[{\citenamefont{Oztekin et~al.}({1995})\citenamefont{Oztekin, Bornside,
  Brown, and Seidel}}]{Oztekin:JApplPhys95}
\bibinfo{author}{\bibfnamefont{A.}~\bibnamefont{Oztekin}},
  \bibinfo{author}{\bibfnamefont{D.}~\bibnamefont{Bornside}},
  \bibinfo{author}{\bibfnamefont{R.}~\bibnamefont{Brown}}, \bibnamefont{and}
  \bibinfo{author}{\bibfnamefont{P.}~\bibnamefont{Seidel}},
  \bibinfo{journal}{{J. Appl. Phys.}} \textbf{\bibinfo{volume}{{77}}},
  \bibinfo{pages}{2297} (\bibinfo{year}{{1995}}).

\bibitem[{\citenamefont{Vanhardeveld et~al.}({1995})\citenamefont{Vanhardeveld,
  Gunter, Vanijzendoorn, Wieldraaijer, Kuipers, and
  Niemantsverdriet}}]{Vanhardeveld:ApplSurfSci95}
\bibinfo{author}{\bibfnamefont{R.}~\bibnamefont{Vanhardeveld}},
  \bibinfo{author}{\bibfnamefont{P.}~\bibnamefont{Gunter}},
  \bibinfo{author}{\bibfnamefont{L.}~\bibnamefont{Vanijzendoorn}},
  \bibinfo{author}{\bibfnamefont{W.}~\bibnamefont{Wieldraaijer}},
  \bibinfo{author}{\bibfnamefont{E.}~\bibnamefont{Kuipers}}, \bibnamefont{and}
  \bibinfo{author}{\bibfnamefont{J.}~\bibnamefont{Niemantsverdriet}},
  \bibinfo{journal}{{Appl. Surf. Sci.}} \textbf{\bibinfo{volume}{{84}}},
  \bibinfo{pages}{339} (\bibinfo{year}{{1995}}).

\bibitem[{\citenamefont{Birnie et~al.}({1995})\citenamefont{Birnie, Zelinski,
  and Perry}}]{Birnie:OptEng95}
\bibinfo{author}{\bibfnamefont{D.}~\bibnamefont{Birnie}},
  \bibinfo{author}{\bibfnamefont{B.}~\bibnamefont{Zelinski}}, \bibnamefont{and}
  \bibinfo{author}{\bibfnamefont{D.}~\bibnamefont{Perry}},
  \bibinfo{journal}{{Opt. Eng.}} \textbf{\bibinfo{volume}{{34}}},
  \bibinfo{pages}{1782} (\bibinfo{year}{{1995}}).

\bibitem[{\citenamefont{Gu et~al.}({1995})\citenamefont{Gu, Bullwinkel, and
  Campbell}}]{Gu:JApplPolSci95}
\bibinfo{author}{\bibfnamefont{J.}~\bibnamefont{Gu}},
  \bibinfo{author}{\bibfnamefont{M.}~\bibnamefont{Bullwinkel}},
  \bibnamefont{and} \bibinfo{author}{\bibfnamefont{G.}~\bibnamefont{Campbell}},
  \bibinfo{journal}{{J. Appl. Polym. Sci.}} \textbf{\bibinfo{volume}{{57}}},
  \bibinfo{pages}{717} (\bibinfo{year}{{1995}}).

\bibitem[{\citenamefont{Gu et~al.}({1996})\citenamefont{Gu, Bullwinkel, and
  Campbell}}]{Gu:PolEngSci96}
\bibinfo{author}{\bibfnamefont{J.}~\bibnamefont{Gu}},
  \bibinfo{author}{\bibfnamefont{M.}~\bibnamefont{Bullwinkel}},
  \bibnamefont{and} \bibinfo{author}{\bibfnamefont{G.}~\bibnamefont{Campbell}},
  \bibinfo{journal}{{Polym. Eng. Sci.}} \textbf{\bibinfo{volume}{{36}}},
  \bibinfo{pages}{1019} (\bibinfo{year}{{1996}}).

\bibitem[{\citenamefont{Birnie and Manley}({1997})}]{Birnie:PhysFluids97}
\bibinfo{author}{\bibfnamefont{D.}~\bibnamefont{Birnie}} \bibnamefont{and}
  \bibinfo{author}{\bibfnamefont{M.}~\bibnamefont{Manley}},
  \bibinfo{journal}{{Phys. Fluids}} \textbf{\bibinfo{volume}{{9}}},
  \bibinfo{pages}{870} (\bibinfo{year}{{1997}}).

\bibitem[{\citenamefont{Birnie}({1997})}]{Birni:SNonCrystSol97}
\bibinfo{author}{\bibfnamefont{D.}~\bibnamefont{Birnie}}, \bibinfo{journal}{{J.
  Non-Cryst. Solids}} \textbf{\bibinfo{volume}{{218}}}, \bibinfo{pages}{174}
  (\bibinfo{year}{{1997}}).

\bibitem[{\citenamefont{Horowitz et~al.}({1998})\citenamefont{Horowitz,
  Michels, and Yeatman}}]{Horowitz:JSolGelSciTech98}
\bibinfo{author}{\bibfnamefont{F.}~\bibnamefont{Horowitz}},
  \bibinfo{author}{\bibfnamefont{A.}~\bibnamefont{Michels}}, \bibnamefont{and}
  \bibinfo{author}{\bibfnamefont{E.}~\bibnamefont{Yeatman}},
  \bibinfo{journal}{{J. Sol-Gel Sci. Technol.}}
  \textbf{\bibinfo{volume}{{13}}}, \bibinfo{pages}{707}
  (\bibinfo{year}{{1998}}).

\bibitem[{\citenamefont{Gupta and Gupta}({1998})}]{Gupta:IndEngChemRes98}
\bibinfo{author}{\bibfnamefont{S.}~\bibnamefont{Gupta}} \bibnamefont{and}
  \bibinfo{author}{\bibfnamefont{R.}~\bibnamefont{Gupta}},
  \bibinfo{journal}{{Ind. Eng. Chem. Res.}} \textbf{\bibinfo{volume}{{37}}},
  \bibinfo{pages}{2223} (\bibinfo{year}{{1998}}).

\bibitem[{\citenamefont{Hall et~al.}({1998})\citenamefont{Hall, Underhill, and
  Torkelson}}]{Hall:PolEngSci98}
\bibinfo{author}{\bibfnamefont{D.}~\bibnamefont{Hall}},
  \bibinfo{author}{\bibfnamefont{P.}~\bibnamefont{Underhill}},
  \bibnamefont{and}
  \bibinfo{author}{\bibfnamefont{J.}~\bibnamefont{Torkelson}},
  \bibinfo{journal}{{Polym. Eng. Sci.}} \textbf{\bibinfo{volume}{{38}}},
  \bibinfo{pages}{2039} (\bibinfo{year}{{1998}}).

\bibitem[{\citenamefont{Haas and
  Quijada}({2000})}]{Haas:ProcSocPhotoOptInstEng00}
\bibinfo{author}{\bibfnamefont{D.}~\bibnamefont{Haas}} \bibnamefont{and}
  \bibinfo{author}{\bibfnamefont{J.}~\bibnamefont{Quijada}}, in
  \emph{\bibinfo{booktitle}{{Sol-Gel Optics V}}}, edited by
  \bibinfo{editor}{\bibnamefont{{Dunn, BS and Pope, EJA and Schmidt, HK and
  Yamane, M}}} (\bibinfo{year}{{2000}}), vol. \bibinfo{volume}{{3943}} of
  \emph{\bibinfo{series}{{Proceedings of the Society of Photo-Optical
  Instrumentation Engineers (SPIE)}}}, pp. \bibinfo{pages}{280--284}.

\bibitem[{\citenamefont{Birnie}({2001})}]{Birnie:JMatRes01}
\bibinfo{author}{\bibfnamefont{D.}~\bibnamefont{Birnie}}, \bibinfo{journal}{{J.
  Mater. Res.}} \textbf{\bibinfo{volume}{{16}}}, \bibinfo{pages}{1145}
  (\bibinfo{year}{{2001}}).

\bibitem[{\citenamefont{Strawhecker et~al.}({2001})\citenamefont{Strawhecker,
  Kumar, Douglas, and Karim}}]{Strawhecker:Macromol01}
\bibinfo{author}{\bibfnamefont{K.}~\bibnamefont{Strawhecker}},
  \bibinfo{author}{\bibfnamefont{S.}~\bibnamefont{Kumar}},
  \bibinfo{author}{\bibfnamefont{J.}~\bibnamefont{Douglas}}, \bibnamefont{and}
  \bibinfo{author}{\bibfnamefont{A.}~\bibnamefont{Karim}},
  \bibinfo{journal}{{Macromolecules}} \textbf{\bibinfo{volume}{{34}}},
  \bibinfo{pages}{4669} (\bibinfo{year}{{2001}}).

\bibitem[{\citenamefont{Walsh and Franses}({2003})}]{Walsh:ThinsolidFilms03}
\bibinfo{author}{\bibfnamefont{C.}~\bibnamefont{Walsh}} \bibnamefont{and}
  \bibinfo{author}{\bibfnamefont{E.}~\bibnamefont{Franses}},
  \bibinfo{journal}{{Thin Solid Films}} \textbf{\bibinfo{volume}{{429}}},
  \bibinfo{pages}{71} (\bibinfo{year}{{2003}}).

\bibitem[{\citenamefont{Burns et~al.}({2003})\citenamefont{Burns, Ramshaw, and
  Jachuck}}]{Burns:ChemEngSci03}
\bibinfo{author}{\bibfnamefont{J.}~\bibnamefont{Burns}},
  \bibinfo{author}{\bibfnamefont{C.}~\bibnamefont{Ramshaw}}, \bibnamefont{and}
  \bibinfo{author}{\bibfnamefont{R.}~\bibnamefont{Jachuck}},
  \bibinfo{journal}{{Chem. Eng. Sci.}} \textbf{\bibinfo{volume}{{58}}},
  \bibinfo{pages}{2245} (\bibinfo{year}{{2003}}).

\bibitem[{\citenamefont{Kim et~al.}({2004})\citenamefont{Kim, Ho, Murphy, and
  Friend}}]{Kim:Macromol04}
\bibinfo{author}{\bibfnamefont{J.}~\bibnamefont{Kim}},
  \bibinfo{author}{\bibfnamefont{P.}~\bibnamefont{Ho}},
  \bibinfo{author}{\bibfnamefont{C.}~\bibnamefont{Murphy}}, \bibnamefont{and}
  \bibinfo{author}{\bibfnamefont{R.}~\bibnamefont{Friend}},
  \bibinfo{journal}{{Macromolecules}} \textbf{\bibinfo{volume}{{37}}},
  \bibinfo{pages}{2861} (\bibinfo{year}{{2004}}).

\bibitem[{\citenamefont{Jukes et~al.}({2005})\citenamefont{Jukes, Heriot,
  Sharp, and Jones}}]{Jukes:Macromol05}
\bibinfo{author}{\bibfnamefont{P.}~\bibnamefont{Jukes}},
  \bibinfo{author}{\bibfnamefont{S.}~\bibnamefont{Heriot}},
  \bibinfo{author}{\bibfnamefont{J.}~\bibnamefont{Sharp}}, \bibnamefont{and}
  \bibinfo{author}{\bibfnamefont{R.}~\bibnamefont{Jones}},
  \bibinfo{journal}{{Macromolecules}} \textbf{\bibinfo{volume}{{38}}},
  \bibinfo{pages}{2030} (\bibinfo{year}{{2005}}).

\bibitem[{\citenamefont{Chang et~al.}({2005})\citenamefont{Chang, Pai, Chen,
  and Jenekhe}}]{Chang:ThinSolidFilms05}
\bibinfo{author}{\bibfnamefont{C.}~\bibnamefont{Chang}},
  \bibinfo{author}{\bibfnamefont{C.}~\bibnamefont{Pai}},
  \bibinfo{author}{\bibfnamefont{W.}~\bibnamefont{Chen}}, \bibnamefont{and}
  \bibinfo{author}{\bibfnamefont{S.}~\bibnamefont{Jenekhe}},
  \bibinfo{journal}{{Thin Solid Films}} \textbf{\bibinfo{volume}{{479}}},
  \bibinfo{pages}{254} (\bibinfo{year}{{2005}}).

\bibitem[{\citenamefont{Birnie et~al.}({2005})\citenamefont{Birnie, Hau,
  Kamber, and Kaz}}]{Birnie:JMatSciElectr05}
\bibinfo{author}{\bibfnamefont{D.}~\bibnamefont{Birnie}},
  \bibinfo{author}{\bibfnamefont{S.}~\bibnamefont{Hau}},
  \bibinfo{author}{\bibfnamefont{D.}~\bibnamefont{Kamber}}, \bibnamefont{and}
  \bibinfo{author}{\bibfnamefont{D.}~\bibnamefont{Kaz}}, \bibinfo{journal}{{J.
  Mater. Sci.-Mater. Electron.}} \textbf{\bibinfo{volume}{{16}}},
  \bibinfo{pages}{715} (\bibinfo{year}{{2005}}).

\bibitem[{\citenamefont{Cheung et~al.}({2005})\citenamefont{Cheung, Grover,
  Wang, Gurkovich, Wang, and Scheinbeim}}]{Cheung:ApplPhysLett05}
\bibinfo{author}{\bibfnamefont{K.}~\bibnamefont{Cheung}},
  \bibinfo{author}{\bibfnamefont{R.}~\bibnamefont{Grover}},
  \bibinfo{author}{\bibfnamefont{Y.}~\bibnamefont{Wang}},
  \bibinfo{author}{\bibfnamefont{C.}~\bibnamefont{Gurkovich}},
  \bibinfo{author}{\bibfnamefont{G.}~\bibnamefont{Wang}}, \bibnamefont{and}
  \bibinfo{author}{\bibfnamefont{J.}~\bibnamefont{Scheinbeim}},
  \bibinfo{journal}{{Appl. Phys. Lett.}} \textbf{\bibinfo{volume}{{87}}}
  (\bibinfo{year}{{2005}}).

\bibitem[{\citenamefont{Yimsiri and Mackley}({2006})}]{Yimsiri:ChemEngSci06}
\bibinfo{author}{\bibfnamefont{P.}~\bibnamefont{Yimsiri}} \bibnamefont{and}
  \bibinfo{author}{\bibfnamefont{M.}~\bibnamefont{Mackley}},
  \bibinfo{journal}{{Chem. Eng. Sci.}} \textbf{\bibinfo{volume}{{61}}},
  \bibinfo{pages}{3496} (\bibinfo{year}{{2006}}).

\bibitem[{\citenamefont{Mohajerani et~al.}({2007})\citenamefont{Mohajerani,
  Farajollahi, Mahzoon, and Baghery}}]{Mohajerani:JOptoAdvMat07}
\bibinfo{author}{\bibfnamefont{E.}~\bibnamefont{Mohajerani}},
  \bibinfo{author}{\bibfnamefont{F.}~\bibnamefont{Farajollahi}},
  \bibinfo{author}{\bibfnamefont{R.}~\bibnamefont{Mahzoon}}, \bibnamefont{and}
  \bibinfo{author}{\bibfnamefont{S.}~\bibnamefont{Baghery}},
  \bibinfo{journal}{{J. Optoelectron. Adv. Mater.}}
  \textbf{\bibinfo{volume}{{9}}}, \bibinfo{pages}{3901}
  (\bibinfo{year}{{2007}}).

\bibitem[{\citenamefont{Parmar and Tirumkudulu}({2009})}]{Parmar:PhysRevE09}
\bibinfo{author}{\bibfnamefont{N.~H.} \bibnamefont{Parmar}} \bibnamefont{and}
  \bibinfo{author}{\bibfnamefont{M.~S.} \bibnamefont{Tirumkudulu}},
  \bibinfo{journal}{{Phys. Rev. E}} \textbf{\bibinfo{volume}{{80}}}
  (\bibinfo{year}{{2009}}).

\bibitem[{\citenamefont{Parmar et~al.}({2009})\citenamefont{Parmar,
  Tirumkudulu, and Hinch}}]{Parmar:PhysFluids09}
\bibinfo{author}{\bibfnamefont{N.~H.} \bibnamefont{Parmar}},
  \bibinfo{author}{\bibfnamefont{M.~S.} \bibnamefont{Tirumkudulu}},
  \bibnamefont{and} \bibinfo{author}{\bibfnamefont{E.~J.} \bibnamefont{Hinch}},
  \bibinfo{journal}{{Phys. Fluids}} \textbf{\bibinfo{volume}{{21}}}
  (\bibinfo{year}{{2009}}).

\bibitem[{\citenamefont{Birnie et~al.}({2010})\citenamefont{Birnie, Haas, and
  Hernandez}}]{Birnie:OptLaserEng10}
\bibinfo{author}{\bibfnamefont{D.~P.} \bibnamefont{Birnie},
  \bibfnamefont{III}}, \bibinfo{author}{\bibfnamefont{D.~E.}
  \bibnamefont{Haas}}, \bibnamefont{and} \bibinfo{author}{\bibfnamefont{C.~M.}
  \bibnamefont{Hernandez}}, \bibinfo{journal}{{Opt. Lasers Eng.}}
  \textbf{\bibinfo{volume}{{48}}}, \bibinfo{pages}{533}
  (\bibinfo{year}{{2010}}).

\bibitem[{\citenamefont{Mokarian-Tabari
  et~al.}({2010})\citenamefont{Mokarian-Tabari, Geoghegan, Howse, Heriot,
  Thompson, and Jones}}]{Mokaraian:EPJE10}
\bibinfo{author}{\bibfnamefont{P.}~\bibnamefont{Mokarian-Tabari}},
  \bibinfo{author}{\bibfnamefont{M.}~\bibnamefont{Geoghegan}},
  \bibinfo{author}{\bibfnamefont{J.~R.} \bibnamefont{Howse}},
  \bibinfo{author}{\bibfnamefont{S.~Y.} \bibnamefont{Heriot}},
  \bibinfo{author}{\bibfnamefont{R.~L.} \bibnamefont{Thompson}},
  \bibnamefont{and} \bibinfo{author}{\bibfnamefont{R.~A.~L.}
  \bibnamefont{Jones}}, \bibinfo{journal}{{Eur. Phys. J. E}}
  \textbf{\bibinfo{volume}{{33}}}, \bibinfo{pages}{283}
  (\bibinfo{year}{{2010}}).

\bibitem[{\citenamefont{Holloway et~al.}({2010})\citenamefont{Holloway,
  Tabuteau, and de~Bruyn}}]{Holloway:RheolActa10}
\bibinfo{author}{\bibfnamefont{K.~E.} \bibnamefont{Holloway}},
  \bibinfo{author}{\bibfnamefont{H.}~\bibnamefont{Tabuteau}}, \bibnamefont{and}
  \bibinfo{author}{\bibfnamefont{J.~R.} \bibnamefont{de~Bruyn}},
  \bibinfo{journal}{{Rheol. Acta}} \textbf{\bibinfo{volume}{{49}}},
  \bibinfo{pages}{245} (\bibinfo{year}{{2010}}).

\bibitem[{\citenamefont{Toolan and Howse}({2013})}]{Toolan:JMatChemC13}
\bibinfo{author}{\bibfnamefont{D.~T.~W.} \bibnamefont{Toolan}}
  \bibnamefont{and} \bibinfo{author}{\bibfnamefont{J.~R.} \bibnamefont{Howse}},
  \bibinfo{journal}{{J. Mater. Chem. C}} \textbf{\bibinfo{volume}{{1}}},
  \bibinfo{pages}{603} (\bibinfo{year}{{2013}}).

\bibitem[{\citenamefont{Acrivos et~al.}(1960)\citenamefont{Acrivos, Shan, and
  Petersen}}]{Acrivos:JApplPhys60}
\bibinfo{author}{\bibfnamefont{A.}~\bibnamefont{Acrivos}},
  \bibinfo{author}{\bibfnamefont{M.}~\bibnamefont{Shan}}, \bibnamefont{and}
  \bibinfo{author}{\bibfnamefont{E.}~\bibnamefont{Petersen}},
  \bibinfo{journal}{J. Appl. Phys.} \textbf{\bibinfo{volume}{31}},
  \bibinfo{pages}{963} (\bibinfo{year}{1960}).

\bibitem[{\citenamefont{Washo}({1977})}]{Washo:IBMJResDev77}
\bibinfo{author}{\bibfnamefont{B.}~\bibnamefont{Washo}}, \bibinfo{journal}{{IBM
  J. Res. Dev.}} \textbf{\bibinfo{volume}{{21}}}, \bibinfo{pages}{190}
  (\bibinfo{year}{{1977}}).

\bibitem[{\citenamefont{Meyerhofer}(1978)}]{Meyerhofer:JApplPhys78}
\bibinfo{author}{\bibfnamefont{D.}~\bibnamefont{Meyerhofer}},
  \bibinfo{journal}{J. Appl. Phys.} \textbf{\bibinfo{volume}{49}},
  \bibinfo{pages}{3993} (\bibinfo{year}{1978}).

\bibitem[{\citenamefont{Matsumoto et~al.}({1982})\citenamefont{Matsumoto,
  Takashima, Kamlya, Kayano, and Ohta}}]{Matsumoto:IndEngChemFundam82}
\bibinfo{author}{\bibfnamefont{S.}~\bibnamefont{Matsumoto}},
  \bibinfo{author}{\bibfnamefont{Y.}~\bibnamefont{Takashima}},
  \bibinfo{author}{\bibfnamefont{T.}~\bibnamefont{Kamlya}},
  \bibinfo{author}{\bibfnamefont{A.}~\bibnamefont{Kayano}}, \bibnamefont{and}
  \bibinfo{author}{\bibfnamefont{Y.}~\bibnamefont{Ohta}},
  \bibinfo{journal}{{Ind. Eng. Chem. Fundam.}} \textbf{\bibinfo{volume}{{21}}},
  \bibinfo{pages}{198} (\bibinfo{year}{{1982}}).

\bibitem[{\citenamefont{Jenekhe}({1983})}]{Jenekhe:PolymEngSci83}
\bibinfo{author}{\bibfnamefont{S.}~\bibnamefont{Jenekhe}},
  \bibinfo{journal}{{Polym. Eng. Sci.}} \textbf{\bibinfo{volume}{{23}}},
  \bibinfo{pages}{830} (\bibinfo{year}{{1983}}).

\bibitem[{\citenamefont{Jenekhe and
  Schuldt}({1984})}]{Jenekhe:IndEngChemFundam84}
\bibinfo{author}{\bibfnamefont{S.}~\bibnamefont{Jenekhe}} \bibnamefont{and}
  \bibinfo{author}{\bibfnamefont{S.}~\bibnamefont{Schuldt}},
  \bibinfo{journal}{{Ind. Eng. Chem. Fundam.}} \textbf{\bibinfo{volume}{{23}}},
  \bibinfo{pages}{432} (\bibinfo{year}{{1984}}).

\bibitem[{\citenamefont{Flack et~al.}(1984)\citenamefont{Flack, Soong, Bell,
  and Hess}}]{Flack:JApplPhys84}
\bibinfo{author}{\bibfnamefont{W.~W.} \bibnamefont{Flack}},
  \bibinfo{author}{\bibfnamefont{D.~S.} \bibnamefont{Soong}},
  \bibinfo{author}{\bibfnamefont{A.~T.} \bibnamefont{Bell}}, \bibnamefont{and}
  \bibinfo{author}{\bibfnamefont{D.~W.} \bibnamefont{Hess}},
  \bibinfo{journal}{J. Appl. Phys.} \textbf{\bibinfo{volume}{56}},
  \bibinfo{pages}{1199} (\bibinfo{year}{1984}).

\bibitem[{\citenamefont{Sukanek}(1985)}]{Sukanek:JImagTech85}
\bibinfo{author}{\bibfnamefont{P.}~\bibnamefont{Sukanek}}, \bibinfo{journal}{J.
  Imag. Technol.} \textbf{\bibinfo{volume}{11}}, \bibinfo{pages}{184}
  (\bibinfo{year}{1985}).

\bibitem[{\citenamefont{Jenekhe and Schuldt}({1985})}]{Jenekhe:ChemEngComm85}
\bibinfo{author}{\bibfnamefont{S.}~\bibnamefont{Jenekhe}} \bibnamefont{and}
  \bibinfo{author}{\bibfnamefont{S.}~\bibnamefont{Schuldt}},
  \bibinfo{journal}{{Chem. Eng. Commun.}} \textbf{\bibinfo{volume}{{33}}},
  \bibinfo{pages}{135} (\bibinfo{year}{{1985}}).

\bibitem[{\citenamefont{Higgins}({1986})}]{Higgins:PhysFluids86}
\bibinfo{author}{\bibfnamefont{B.}~\bibnamefont{Higgins}},
  \bibinfo{journal}{{Phys. Fluids}} \textbf{\bibinfo{volume}{{29}}},
  \bibinfo{pages}{3522} (\bibinfo{year}{{1986}}).

\bibitem[{\citenamefont{Kaplon et~al.}(1986)\citenamefont{Kaplon, Kawala, and
  Skoczylas}}]{Kaplon:ChemEngSci86}
\bibinfo{author}{\bibfnamefont{J.}~\bibnamefont{Kaplon}},
  \bibinfo{author}{\bibfnamefont{Z.}~\bibnamefont{Kawala}}, \bibnamefont{and}
  \bibinfo{author}{\bibfnamefont{A.}~\bibnamefont{Skoczylas}},
  \bibinfo{journal}{Chem. Eng. Sci.} \textbf{\bibinfo{volume}{41}},
  \bibinfo{pages}{519} (\bibinfo{year}{1986}).

\bibitem[{\citenamefont{Bornside et~al.}(1987)\citenamefont{Bornside, Macosko,
  and Scriven}}]{Bornside:JImagTech87}
\bibinfo{author}{\bibfnamefont{D.}~\bibnamefont{Bornside}},
  \bibinfo{author}{\bibfnamefont{C.}~\bibnamefont{Macosko}}, \bibnamefont{and}
  \bibinfo{author}{\bibfnamefont{L.}~\bibnamefont{Scriven}},
  \bibinfo{journal}{J. Imag. Techn.} \textbf{\bibinfo{volume}{13}},
  \bibinfo{pages}{122} (\bibinfo{year}{1987}).

\bibitem[{\citenamefont{Shimoji}(1987)}]{Shimoji:JapanJApplPhys87}
\bibinfo{author}{\bibfnamefont{S.}~\bibnamefont{Shimoji}},
  \bibinfo{journal}{Japan. J. Appl. Phys.} \textbf{\bibinfo{volume}{26}},
  \bibinfo{pages}{L905} (\bibinfo{year}{1987}).

\bibitem[{\citenamefont{Yanagisawa}({1987})}]{Yanagisawa:JApplPhys87}
\bibinfo{author}{\bibfnamefont{M.}~\bibnamefont{Yanagisawa}},
  \bibinfo{journal}{{J. Appl. Phys.}} \textbf{\bibinfo{volume}{{61}}},
  \bibinfo{pages}{1034} (\bibinfo{year}{{1987}}).

\bibitem[{\citenamefont{Tu}({1987})}]{Tu:JCollInterfSci87}
\bibinfo{author}{\bibfnamefont{Y.}~\bibnamefont{Tu}}, \bibinfo{journal}{{J.
  Colloid Interface Sci.}} \textbf{\bibinfo{volume}{{116}}},
  \bibinfo{pages}{237} (\bibinfo{year}{{1987}}).

\bibitem[{\citenamefont{Stillwagon et~al.}({1987})\citenamefont{Stillwagon,
  Larson, and Taylor}}]{Stillwagon:JElChemSoc87}
\bibinfo{author}{\bibfnamefont{L.}~\bibnamefont{Stillwagon}},
  \bibinfo{author}{\bibfnamefont{R.}~\bibnamefont{Larson}}, \bibnamefont{and}
  \bibinfo{author}{\bibfnamefont{G.}~\bibnamefont{Taylor}},
  \bibinfo{journal}{{J. Electrochem. Soc.}} \textbf{\bibinfo{volume}{{134}}},
  \bibinfo{pages}{2030} (\bibinfo{year}{{1987}}).

\bibitem[{\citenamefont{Middleman}({1987})}]{Middleman:JApplPhys87}
\bibinfo{author}{\bibfnamefont{S.}~\bibnamefont{Middleman}},
  \bibinfo{journal}{{J. Appl. Phys.}} \textbf{\bibinfo{volume}{{62}}},
  \bibinfo{pages}{2530} (\bibinfo{year}{{1987}}).

\bibitem[{\citenamefont{Rehg and Higgins}({1988})}]{Rehg:PhysFluids88}
\bibinfo{author}{\bibfnamefont{T.}~\bibnamefont{Rehg}} \bibnamefont{and}
  \bibinfo{author}{\bibfnamefont{B.}~\bibnamefont{Higgins}},
  \bibinfo{journal}{{Phys. Fluids}} \textbf{\bibinfo{volume}{{31}}},
  \bibinfo{pages}{1360} (\bibinfo{year}{{1988}}).

\bibitem[{\citenamefont{Papanastasiou
  et~al.}({1988})\citenamefont{Papanastasiou, Alexandrou, and
  Graebel}}]{Papanstasiou:JRehol88}
\bibinfo{author}{\bibfnamefont{T.}~\bibnamefont{Papanastasiou}},
  \bibinfo{author}{\bibfnamefont{A.}~\bibnamefont{Alexandrou}},
  \bibnamefont{and} \bibinfo{author}{\bibfnamefont{W.}~\bibnamefont{Graebel}},
  \bibinfo{journal}{{J. Rheol.}} \textbf{\bibinfo{volume}{{32}}},
  \bibinfo{pages}{485} (\bibinfo{year}{{1988}}).

\bibitem[{\citenamefont{Lawrece}(1988)}]{Lawrence:PhysFluids88}
\bibinfo{author}{\bibfnamefont{C.}~\bibnamefont{Lawrece}},
  \bibinfo{journal}{Phys. Fluids} \textbf{\bibinfo{volume}{31}},
  \bibinfo{pages}{2786} (\bibinfo{year}{1988}).

\bibitem[{\citenamefont{Bornside et~al.}(1989)\citenamefont{Bornside, Macosko,
  and Scriven}}]{Bornside:JApplPhys89}
\bibinfo{author}{\bibfnamefont{D.}~\bibnamefont{Bornside}},
  \bibinfo{author}{\bibfnamefont{C.}~\bibnamefont{Macosko}}, \bibnamefont{and}
  \bibinfo{author}{\bibfnamefont{L.}~\bibnamefont{Scriven}},
  \bibinfo{journal}{J. Appl. Phys.} \textbf{\bibinfo{volume}{66}},
  \bibinfo{pages}{5185} (\bibinfo{year}{1989}).

\bibitem[{\citenamefont{Ohara et~al.}(1989)\citenamefont{Ohara, Matsumoto, and
  Ohashi}}]{Ohara:PhysFluidsA89}
\bibinfo{author}{\bibfnamefont{T.}~\bibnamefont{Ohara}},
  \bibinfo{author}{\bibfnamefont{Y.}~\bibnamefont{Matsumoto}},
  \bibnamefont{and} \bibinfo{author}{\bibfnamefont{H.}~\bibnamefont{Ohashi}},
  \bibinfo{journal}{Phys. Fluids A} \textbf{\bibinfo{volume}{1}},
  \bibinfo{pages}{1949} (\bibinfo{year}{1989}).

\bibitem[{\citenamefont{Shimoji}(1989)}]{Shimoji:JApplPhys89}
\bibinfo{author}{\bibfnamefont{S.}~\bibnamefont{Shimoji}}, \bibinfo{journal}{J.
  Appl. Phys.} \textbf{\bibinfo{volume}{66}}, \bibinfo{pages}{2712}
  (\bibinfo{year}{1989}).

\bibitem[{\citenamefont{Matsumoto et~al.}({1989})\citenamefont{Matsumoto,
  Ohara, Teruya, and Ohashi}}]{Matsumoto:JSMEIntJ89}
\bibinfo{author}{\bibfnamefont{Y.}~\bibnamefont{Matsumoto}},
  \bibinfo{author}{\bibfnamefont{T.}~\bibnamefont{Ohara}},
  \bibinfo{author}{\bibfnamefont{I.}~\bibnamefont{Teruya}}, \bibnamefont{and}
  \bibinfo{author}{\bibfnamefont{H.}~\bibnamefont{Ohashi}},
  \bibinfo{journal}{{JSME Int. J.}} \textbf{\bibinfo{volume}{{32}}},
  \bibinfo{pages}{52} (\bibinfo{year}{{1989}}).

\bibitem[{\citenamefont{Hwang and Ma}({1989})}]{Hwang:JApplPhys89}
\bibinfo{author}{\bibfnamefont{J.}~\bibnamefont{Hwang}} \bibnamefont{and}
  \bibinfo{author}{\bibfnamefont{F.}~\bibnamefont{Ma}}, \bibinfo{journal}{{J.
  Appl. Phys.}} \textbf{\bibinfo{volume}{{66}}}, \bibinfo{pages}{388}
  (\bibinfo{year}{{1989}}).

\bibitem[{\citenamefont{Ma and Hwang}({1989})}]{Ma:JApplPhys89}
\bibinfo{author}{\bibfnamefont{F.}~\bibnamefont{Ma}} \bibnamefont{and}
  \bibinfo{author}{\bibfnamefont{J.}~\bibnamefont{Hwang}},
  \bibinfo{journal}{{J. Appl. Phys.}} \textbf{\bibinfo{volume}{{66}}},
  \bibinfo{pages}{5026} (\bibinfo{year}{{1989}}).

\bibitem[{\citenamefont{Dandapat and Ray}({1990})}]{Dandapat:IntJNonlinMech90}
\bibinfo{author}{\bibfnamefont{B.}~\bibnamefont{Dandapat}} \bibnamefont{and}
  \bibinfo{author}{\bibfnamefont{P.}~\bibnamefont{Ray}},
  \bibinfo{journal}{{Int. J. Non-Linear Mech.}}
  \textbf{\bibinfo{volume}{{25}}}, \bibinfo{pages}{569}
  (\bibinfo{year}{{1990}}).

\bibitem[{\citenamefont{Ma and Hwang}({1990})}]{Ma:JApplPhys90}
\bibinfo{author}{\bibfnamefont{F.}~\bibnamefont{Ma}} \bibnamefont{and}
  \bibinfo{author}{\bibfnamefont{J.}~\bibnamefont{Hwang}},
  \bibinfo{journal}{{J. Appl. Phys.}} \textbf{\bibinfo{volume}{{68}}},
  \bibinfo{pages}{1265} (\bibinfo{year}{{1990}}).

\bibitem[{\citenamefont{Stillwagon and
  Larson}({1990})}]{Stillwagon:PhysFluidsA90}
\bibinfo{author}{\bibfnamefont{L.}~\bibnamefont{Stillwagon}} \bibnamefont{and}
  \bibinfo{author}{\bibfnamefont{R.}~\bibnamefont{Larson}},
  \bibinfo{journal}{Phys. Fluids A} \textbf{\bibinfo{volume}{{2}}},
  \bibinfo{pages}{1937} (\bibinfo{year}{{1990}}).

\bibitem[{\citenamefont{Hwang and Ma}({1990})}]{Hwang:MechResComm90}
\bibinfo{author}{\bibfnamefont{J.}~\bibnamefont{Hwang}} \bibnamefont{and}
  \bibinfo{author}{\bibfnamefont{F.}~\bibnamefont{Ma}},
  \bibinfo{journal}{{Mech. Res. Commun.}} \textbf{\bibinfo{volume}{{17}}},
  \bibinfo{pages}{423} (\bibinfo{year}{{1990}}).

\bibitem[{\citenamefont{Lawrence}(1990)}]{Lawrence:PhysFluidsA90}
\bibinfo{author}{\bibfnamefont{C.}~\bibnamefont{Lawrence}},
  \bibinfo{journal}{Phys. Fluids A} \textbf{\bibinfo{volume}{2}},
  \bibinfo{pages}{453} (\bibinfo{year}{1990}).

\bibitem[{\citenamefont{Tu and Drake}(1990)}]{Tu:JCollInterfSci90}
\bibinfo{author}{\bibfnamefont{Y.}~\bibnamefont{Tu}} \bibnamefont{and}
  \bibinfo{author}{\bibfnamefont{R.}~\bibnamefont{Drake}}, \bibinfo{journal}{J.
  Coll. Interf. Sci} \textbf{\bibinfo{volume}{135}}, \bibinfo{pages}{562}
  (\bibinfo{year}{1990}).

\bibitem[{\citenamefont{Lawrence and
  Zhou}(1991)}]{Lawrence:JNonNewtonianFluidMech91}
\bibinfo{author}{\bibfnamefont{C.}~\bibnamefont{Lawrence}} \bibnamefont{and}
  \bibinfo{author}{\bibfnamefont{W.}~\bibnamefont{Zhou}}, \bibinfo{journal}{J.
  Non Newtonian Fluid Mech.} \textbf{\bibinfo{volume}{39}},
  \bibinfo{pages}{137} (\bibinfo{year}{1991}).

\bibitem[{\citenamefont{Reisfeld
  et~al.}(1991{\natexlab{a}})\citenamefont{Reisfeld, Bankoff, and
  Davis}}]{Reisfeld:JApplPhys91:1}
\bibinfo{author}{\bibfnamefont{B.}~\bibnamefont{Reisfeld}},
  \bibinfo{author}{\bibfnamefont{S.}~\bibnamefont{Bankoff}}, \bibnamefont{and}
  \bibinfo{author}{\bibfnamefont{S.}~\bibnamefont{Davis}}, \bibinfo{journal}{J.
  Appl. Phys.} \textbf{\bibinfo{volume}{70}}, \bibinfo{pages}{5258}
  (\bibinfo{year}{1991}{\natexlab{a}}).

\bibitem[{\citenamefont{Reisfeld
  et~al.}(1991{\natexlab{b}})\citenamefont{Reisfeld, Bankoff, and
  Davis}}]{Reisfeld:JApplPhys91:2}
\bibinfo{author}{\bibfnamefont{B.}~\bibnamefont{Reisfeld}},
  \bibinfo{author}{\bibfnamefont{S.}~\bibnamefont{Bankoff}}, \bibnamefont{and}
  \bibinfo{author}{\bibfnamefont{S.}~\bibnamefont{Davis}}, \bibinfo{journal}{J.
  Appl. Phys.} \textbf{\bibinfo{volume}{70}}, \bibinfo{pages}{5267}
  (\bibinfo{year}{1991}{\natexlab{b}}).

\bibitem[{\citenamefont{Sukanek}(1991)}]{Sukanek:JElChemSoc91}
\bibinfo{author}{\bibfnamefont{P.}~\bibnamefont{Sukanek}}, \bibinfo{journal}{J.
  Electrochem. Soc.} \textbf{\bibinfo{volume}{138}}, \bibinfo{pages}{1712}
  (\bibinfo{year}{1991}).

\bibitem[{\citenamefont{Wang et~al.}({1991})\citenamefont{Wang, Watson, and
  Alexander}}]{Wang:IMAJApplMath91}
\bibinfo{author}{\bibfnamefont{C.}~\bibnamefont{Wang}},
  \bibinfo{author}{\bibfnamefont{L.}~\bibnamefont{Watson}}, \bibnamefont{and}
  \bibinfo{author}{\bibfnamefont{K.}~\bibnamefont{Alexander}},
  \bibinfo{journal}{{IMA J. Appl. Math.}} \textbf{\bibinfo{volume}{{46}}},
  \bibinfo{pages}{201} (\bibinfo{year}{{1991}}).

\bibitem[{\citenamefont{Kim et~al.}({1991})\citenamefont{Kim, Kim, and
  Ma}}]{Kim:JApplPhys91}
\bibinfo{author}{\bibfnamefont{S.}~\bibnamefont{Kim}},
  \bibinfo{author}{\bibfnamefont{J.}~\bibnamefont{Kim}}, \bibnamefont{and}
  \bibinfo{author}{\bibfnamefont{F.}~\bibnamefont{Ma}}, \bibinfo{journal}{{J.
  Appl. Phys.}} \textbf{\bibinfo{volume}{{69}}}, \bibinfo{pages}{2593}
  (\bibinfo{year}{{1991}}).

\bibitem[{\citenamefont{Potanin}(1992)}]{Potanin:ChemEngSci92}
\bibinfo{author}{\bibfnamefont{A.}~\bibnamefont{Potanin}},
  \bibinfo{journal}{Chem. Eng. Sci.} \textbf{\bibinfo{volume}{47}},
  \bibinfo{pages}{1871} (\bibinfo{year}{1992}).

\bibitem[{\citenamefont{Stillwagon and
  Larson}(1992)}]{Stillwagon:PhysFluidsA92}
\bibinfo{author}{\bibfnamefont{L.}~\bibnamefont{Stillwagon}} \bibnamefont{and}
  \bibinfo{author}{\bibfnamefont{R.}~\bibnamefont{Larson}},
  \bibinfo{journal}{Phys. Fluids A} \textbf{\bibinfo{volume}{4}},
  \bibinfo{pages}{895} (\bibinfo{year}{1992}).

\bibitem[{\citenamefont{Yonkoski and Soane}(1992)}]{Yonkoski:JApplPhys92}
\bibinfo{author}{\bibfnamefont{R.}~\bibnamefont{Yonkoski}} \bibnamefont{and}
  \bibinfo{author}{\bibfnamefont{D.}~\bibnamefont{Soane}}, \bibinfo{journal}{J.
  Appl. Phys.} \textbf{\bibinfo{volume}{72}}, \bibinfo{pages}{725}
  (\bibinfo{year}{1992}).

\bibitem[{\citenamefont{Rehg and Higgins}(1992)}]{Rehg:AIChEJ92}
\bibinfo{author}{\bibfnamefont{T.}~\bibnamefont{Rehg}} \bibnamefont{and}
  \bibinfo{author}{\bibfnamefont{B.}~\bibnamefont{Higgins}},
  \bibinfo{journal}{AIChE J.} \textbf{\bibinfo{volume}{38}},
  \bibinfo{pages}{489} (\bibinfo{year}{1992}).

\bibitem[{\citenamefont{Bornside et~al.}(1993)\citenamefont{Bornside, Brown,
  Ackmann, Frank, Tryba, and Geyling}}]{Bornside:JApplPhys93}
\bibinfo{author}{\bibfnamefont{D.}~\bibnamefont{Bornside}},
  \bibinfo{author}{\bibfnamefont{R.}~\bibnamefont{Brown}},
  \bibinfo{author}{\bibfnamefont{P.}~\bibnamefont{Ackmann}},
  \bibinfo{author}{\bibfnamefont{J.}~\bibnamefont{Frank}},
  \bibinfo{author}{\bibfnamefont{A.}~\bibnamefont{Tryba}}, \bibnamefont{and}
  \bibinfo{author}{\bibfnamefont{F.}~\bibnamefont{Geyling}},
  \bibinfo{journal}{J. Appl. Phys.} \textbf{\bibinfo{volume}{73}},
  \bibinfo{pages}{585} (\bibinfo{year}{1993}).

\bibitem[{\citenamefont{Levinson et~al.}(1993)\citenamefont{Levinson, Arnold,
  and Dehodgins}}]{Levinson:PolEngSci93}
\bibinfo{author}{\bibfnamefont{W.}~\bibnamefont{Levinson}},
  \bibinfo{author}{\bibfnamefont{A.}~\bibnamefont{Arnold}}, \bibnamefont{and}
  \bibinfo{author}{\bibfnamefont{O.}~\bibnamefont{Dehodgins}},
  \bibinfo{journal}{Pol. Eng. Sci.} \textbf{\bibinfo{volume}{33}},
  \bibinfo{pages}{980} (\bibinfo{year}{1993}).

\bibitem[{\citenamefont{Dandapat and Ray}({1993})}]{Dandapat:IntJNonlinMech93}
\bibinfo{author}{\bibfnamefont{B.}~\bibnamefont{Dandapat}} \bibnamefont{and}
  \bibinfo{author}{\bibfnamefont{P.}~\bibnamefont{Ray}},
  \bibinfo{journal}{{Int. J. Non-Linear Mech.}}
  \textbf{\bibinfo{volume}{{28}}}, \bibinfo{pages}{489}
  (\bibinfo{year}{{1993}}).

\bibitem[{\citenamefont{Kim et~al.}({1993})\citenamefont{Kim, Kim, and
  Ma}}]{Kim:JApplPhys93}
\bibinfo{author}{\bibfnamefont{J.}~\bibnamefont{Kim}},
  \bibinfo{author}{\bibfnamefont{S.}~\bibnamefont{Kim}}, \bibnamefont{and}
  \bibinfo{author}{\bibfnamefont{F.}~\bibnamefont{Ma}}, \bibinfo{journal}{{J.
  Appl. Phys.}} \textbf{\bibinfo{volume}{{73}}}, \bibinfo{pages}{422}
  (\bibinfo{year}{{1993}}).

\bibitem[{\citenamefont{Forcada and Mate}({1993})}]{Forcada:JCollInterfSci93}
\bibinfo{author}{\bibfnamefont{M.}~\bibnamefont{Forcada}} \bibnamefont{and}
  \bibinfo{author}{\bibfnamefont{C.}~\bibnamefont{Mate}}, \bibinfo{journal}{{J.
  Colloid Interface Sci.}} \textbf{\bibinfo{volume}{{160}}},
  \bibinfo{pages}{218} (\bibinfo{year}{{1993}}).

\bibitem[{\citenamefont{Ma}({1994})}]{Ma:ProbabEngEngMech94}
\bibinfo{author}{\bibfnamefont{F.}~\bibnamefont{Ma}},
  \bibinfo{journal}{{Probab. Eng. Eng. Mech.}} \textbf{\bibinfo{volume}{{9}}},
  \bibinfo{pages}{39} (\bibinfo{year}{{1994}}).

\bibitem[{\citenamefont{Spaid and Homsy}({1994})}]{Spaid:JNonnewtonFluidMech94}
\bibinfo{author}{\bibfnamefont{M.}~\bibnamefont{Spaid}} \bibnamefont{and}
  \bibinfo{author}{\bibfnamefont{G.}~\bibnamefont{Homsy}},
  \bibinfo{journal}{{J. Non-Newton. Fluid Mech.}}
  \textbf{\bibinfo{volume}{{55}}}, \bibinfo{pages}{249}
  (\bibinfo{year}{{1994}}).

\bibitem[{\citenamefont{Borkar et~al.}(1994)\citenamefont{Borkar, Tsamopoulos,
  and Gupta}}]{Borkar:PhysFluids94}
\bibinfo{author}{\bibfnamefont{A.}~\bibnamefont{Borkar}},
  \bibinfo{author}{\bibfnamefont{J.}~\bibnamefont{Tsamopoulos}},
  \bibnamefont{and} \bibinfo{author}{\bibfnamefont{R.}~\bibnamefont{Gupta}},
  \bibinfo{journal}{Phys. Fluids} \textbf{\bibinfo{volume}{6}},
  \bibinfo{pages}{3539} (\bibinfo{year}{1994}).

\bibitem[{\citenamefont{Dandapat and Ray}({1994})}]{Dandapat:JPhysD94}
\bibinfo{author}{\bibfnamefont{B.}~\bibnamefont{Dandapat}} \bibnamefont{and}
  \bibinfo{author}{\bibfnamefont{P.}~\bibnamefont{Ray}}, \bibinfo{journal}{{J.
  Phys. D-Appl. Phys.}} \textbf{\bibinfo{volume}{{27}}}, \bibinfo{pages}{2041}
  (\bibinfo{year}{{1994}}).

\bibitem[{\citenamefont{Gu et~al.}(1995)\citenamefont{Gu, Bullwinkel, and
  Campbell}}]{Gu:JElectrochemSoc95}
\bibinfo{author}{\bibfnamefont{J.}~\bibnamefont{Gu}},
  \bibinfo{author}{\bibfnamefont{M.~D.} \bibnamefont{Bullwinkel}},
  \bibnamefont{and} \bibinfo{author}{\bibfnamefont{G.~A.}
  \bibnamefont{Campbell}}, \bibinfo{journal}{J. Electrochem. Soc.}
  \textbf{\bibinfo{volume}{142}}, \bibinfo{pages}{907} (\bibinfo{year}{1995}).

\bibitem[{\citenamefont{da~Souza et~al.}(1995)\citenamefont{da~Souza, Leaver,
  and Eskiyerli}}]{deSouza:CompMatSci95}
\bibinfo{author}{\bibfnamefont{M.}~\bibnamefont{da~Souza}},
  \bibinfo{author}{\bibfnamefont{K.}~\bibnamefont{Leaver}}, \bibnamefont{and}
  \bibinfo{author}{\bibfnamefont{M.}~\bibnamefont{Eskiyerli}},
  \bibinfo{journal}{Comput. Mat. Sci.} \textbf{\bibinfo{volume}{4}},
  \bibinfo{pages}{233} (\bibinfo{year}{1995}).

\bibitem[{\citenamefont{Wnag and Yen}(1995)}]{Wang:ChemEngSci95}
\bibinfo{author}{\bibfnamefont{C.-T.} \bibnamefont{Wnag}} \bibnamefont{and}
  \bibinfo{author}{\bibfnamefont{S.-C.} \bibnamefont{Yen}},
  \bibinfo{journal}{Chem. Eng. Sci.} \textbf{\bibinfo{volume}{50}},
  \bibinfo{pages}{989} (\bibinfo{year}{1995}).

\bibitem[{\citenamefont{Yen and Wnag}(1995)}]{Yen:JChinIChE95}
\bibinfo{author}{\bibfnamefont{S.-C.} \bibnamefont{Yen}} \bibnamefont{and}
  \bibinfo{author}{\bibfnamefont{C.-T.} \bibnamefont{Wnag}},
  \bibinfo{journal}{J. Chin. I. Ch. E.} \textbf{\bibinfo{volume}{26}},
  \bibinfo{pages}{157} (\bibinfo{year}{1995}).

\bibitem[{\citenamefont{van Hardeveld et~al.}(1995)\citenamefont{van Hardeveld,
  van IJzendoorn, Wieldraaijer, Kuipers, and
  Niemantsverdriet}}]{Hardeveld:AppsSurfSci95}
\bibinfo{author}{\bibfnamefont{R.}~\bibnamefont{van Hardeveld}},
  \bibinfo{author}{\bibfnamefont{P.~G.~L.} \bibnamefont{van IJzendoorn}},
  \bibinfo{author}{\bibfnamefont{W.}~\bibnamefont{Wieldraaijer}},
  \bibinfo{author}{\bibfnamefont{E.}~\bibnamefont{Kuipers}}, \bibnamefont{and}
  \bibinfo{author}{\bibfnamefont{J.}~\bibnamefont{Niemantsverdriet}},
  \bibinfo{journal}{Appl. Surf. Sci.} \textbf{\bibinfo{volume}{84}},
  \bibinfo{pages}{339} (\bibinfo{year}{1995}).

\bibitem[{\citenamefont{Burgess and Wilson}(1996)}]{Burgess:PhysFluids96}
\bibinfo{author}{\bibfnamefont{S.}~\bibnamefont{Burgess}} \bibnamefont{and}
  \bibinfo{author}{\bibfnamefont{S.}~\bibnamefont{Wilson}},
  \bibinfo{journal}{Phys. Fluids} \textbf{\bibinfo{volume}{8}},
  \bibinfo{pages}{2291} (\bibinfo{year}{1996}).

\bibitem[{\citenamefont{Tsamopoulos et~al.}({1996})\citenamefont{Tsamopoulos,
  Chen, and Borkar}}]{Tsamopoulos:RheolActa96}
\bibinfo{author}{\bibfnamefont{J.}~\bibnamefont{Tsamopoulos}},
  \bibinfo{author}{\bibfnamefont{M.}~\bibnamefont{Chen}}, \bibnamefont{and}
  \bibinfo{author}{\bibfnamefont{A.}~\bibnamefont{Borkar}},
  \bibinfo{journal}{{Rheol. Acta}} \textbf{\bibinfo{volume}{{35}}},
  \bibinfo{pages}{597} (\bibinfo{year}{{1996}}).

\bibitem[{\citenamefont{Okuzono et~al.}(2006)\citenamefont{Okuzono, Ozawa, and
  Doi}}]{Okuzono:PhysRevLett97}
\bibinfo{author}{\bibfnamefont{T.}~\bibnamefont{Okuzono}},
  \bibinfo{author}{\bibfnamefont{K.}~\bibnamefont{Ozawa}}, \bibnamefont{and}
  \bibinfo{author}{\bibfnamefont{M.}~\bibnamefont{Doi}},
  \bibinfo{journal}{Phys. Rev. Lett.} \textbf{\bibinfo{volume}{97}},
  \bibinfo{pages}{136103} (\bibinfo{year}{2006}).

\bibitem[{\citenamefont{Sukanek}(1997)}]{Sukanek:JElChemSoc97}
\bibinfo{author}{\bibfnamefont{P.}~\bibnamefont{Sukanek}}, \bibinfo{journal}{J.
  Electrochem. Soc.} \textbf{\bibinfo{volume}{144}}, \bibinfo{pages}{3959}
  (\bibinfo{year}{1997}).

\bibitem[{\citenamefont{Momoniat and
  Mason}({1998})}]{Momoniat:IntJNonlinMech98}
\bibinfo{author}{\bibfnamefont{E.}~\bibnamefont{Momoniat}} \bibnamefont{and}
  \bibinfo{author}{\bibfnamefont{D.}~\bibnamefont{Mason}},
  \bibinfo{journal}{{Int. J. Non-Linear Mech.}}
  \textbf{\bibinfo{volume}{{33}}}, \bibinfo{pages}{1069}
  (\bibinfo{year}{{1998}}).

\bibitem[{\citenamefont{Dandapat and Ray}({1998})}]{Dandapat:ZAngewMathMech98}
\bibinfo{author}{\bibfnamefont{B.}~\bibnamefont{Dandapat}} \bibnamefont{and}
  \bibinfo{author}{\bibfnamefont{P.}~\bibnamefont{Ray}}, \bibinfo{journal}{{Z.
  Angew. Math. Mech.}} \textbf{\bibinfo{volume}{{78}}}, \bibinfo{pages}{635}
  (\bibinfo{year}{{1998}}).

\bibitem[{\citenamefont{Dandapat and Layek}({1999})}]{Dandapat:JPhysD99}
\bibinfo{author}{\bibfnamefont{B.}~\bibnamefont{Dandapat}} \bibnamefont{and}
  \bibinfo{author}{\bibfnamefont{G.}~\bibnamefont{Layek}},
  \bibinfo{journal}{{J. Phys. D-Appl. Phys.}} \textbf{\bibinfo{volume}{{32}}},
  \bibinfo{pages}{2483} (\bibinfo{year}{{1999}}).

\bibitem[{\citenamefont{McKinley et~al.}({1999})\citenamefont{McKinley, Wilson,
  and Duffy}}]{McKinley:PhysFluids99}
\bibinfo{author}{\bibfnamefont{I.}~\bibnamefont{McKinley}},
  \bibinfo{author}{\bibfnamefont{S.}~\bibnamefont{Wilson}}, \bibnamefont{and}
  \bibinfo{author}{\bibfnamefont{B.}~\bibnamefont{Duffy}},
  \bibinfo{journal}{{Phys. Fluids}} \textbf{\bibinfo{volume}{{11}}},
  \bibinfo{pages}{30} (\bibinfo{year}{{1999}}).

\bibitem[{\citenamefont{Wu and Chou}({1999})}]{Wu:JElChemSoc99}
\bibinfo{author}{\bibfnamefont{P.}~\bibnamefont{Wu}} \bibnamefont{and}
  \bibinfo{author}{\bibfnamefont{F.}~\bibnamefont{Chou}}, \bibinfo{journal}{{J.
  Electrochem. Soc.}} \textbf{\bibinfo{volume}{{146}}}, \bibinfo{pages}{3819}
  (\bibinfo{year}{{1999}}).

\bibitem[{\citenamefont{Wilson et~al.}({2000})\citenamefont{Wilson, Hunt, and
  Duffy}}]{Wilson:JFluidMech00}
\bibinfo{author}{\bibfnamefont{S.}~\bibnamefont{Wilson}},
  \bibinfo{author}{\bibfnamefont{R.}~\bibnamefont{Hunt}}, \bibnamefont{and}
  \bibinfo{author}{\bibfnamefont{B.}~\bibnamefont{Duffy}},
  \bibinfo{journal}{{J. Fluid Mech.}} \textbf{\bibinfo{volume}{{413}}},
  \bibinfo{pages}{65} (\bibinfo{year}{{2000}}).

\bibitem[{\citenamefont{Kitamura}({2000})}]{Kitamura:PhysFluids00}
\bibinfo{author}{\bibfnamefont{A.}~\bibnamefont{Kitamura}},
  \bibinfo{journal}{{Phys. Fluids}} \textbf{\bibinfo{volume}{{12}}},
  \bibinfo{pages}{2141} (\bibinfo{year}{{2000}}).

\bibitem[{\citenamefont{Chou and Wu}({2000})}]{Chou:JElChemSoc00}
\bibinfo{author}{\bibfnamefont{F.}~\bibnamefont{Chou}} \bibnamefont{and}
  \bibinfo{author}{\bibfnamefont{P.}~\bibnamefont{Wu}}, \bibinfo{journal}{{J.
  Electrochem. Soc.}} \textbf{\bibinfo{volume}{{147}}}, \bibinfo{pages}{699}
  (\bibinfo{year}{{2000}}).

\bibitem[{\citenamefont{Dandapat}({2001})}]{Dandapat:PhysFluids01}
\bibinfo{author}{\bibfnamefont{B.}~\bibnamefont{Dandapat}},
  \bibinfo{journal}{{Phys. Fluids}} \textbf{\bibinfo{volume}{{13}}},
  \bibinfo{pages}{1860} (\bibinfo{year}{{2001}}).

\bibitem[{\citenamefont{Usha and Ravindran}({2001})}]{Usha:IntJNonlinMech01}
\bibinfo{author}{\bibfnamefont{R.}~\bibnamefont{Usha}} \bibnamefont{and}
  \bibinfo{author}{\bibfnamefont{R.}~\bibnamefont{Ravindran}},
  \bibinfo{journal}{{Int. J. Non-Linear Mech.}}
  \textbf{\bibinfo{volume}{{36}}}, \bibinfo{pages}{147}
  (\bibinfo{year}{{2001}}).

\bibitem[{\citenamefont{Usha and Gotz}({2001})}]{Usha:ActaMech01}
\bibinfo{author}{\bibfnamefont{R.}~\bibnamefont{Usha}} \bibnamefont{and}
  \bibinfo{author}{\bibfnamefont{T.}~\bibnamefont{Gotz}},
  \bibinfo{journal}{{Acta Mech.}} \textbf{\bibinfo{volume}{{147}}},
  \bibinfo{pages}{137} (\bibinfo{year}{{2001}}).

\bibitem[{\citenamefont{Myers and Charpin}({2001})}]{Myres:IntJNonlinMech01}
\bibinfo{author}{\bibfnamefont{T.}~\bibnamefont{Myers}} \bibnamefont{and}
  \bibinfo{author}{\bibfnamefont{J.}~\bibnamefont{Charpin}},
  \bibinfo{journal}{{Int. J. Non-Linear Mech.}}
  \textbf{\bibinfo{volume}{{36}}}, \bibinfo{pages}{629}
  (\bibinfo{year}{{2001}}).

\bibitem[{\citenamefont{Usha and Uma}({2001})}]{Usha:ZAngewMathPhys01}
\bibinfo{author}{\bibfnamefont{R.}~\bibnamefont{Usha}} \bibnamefont{and}
  \bibinfo{author}{\bibfnamefont{B.}~\bibnamefont{Uma}}, \bibinfo{journal}{{Z.
  Angew. Math. Phys.}} \textbf{\bibinfo{volume}{{52}}}, \bibinfo{pages}{793}
  (\bibinfo{year}{{2001}}).

\bibitem[{\citenamefont{Kitamura}({2001})}]{Kitamura:PhysFluids01}
\bibinfo{author}{\bibfnamefont{A.}~\bibnamefont{Kitamura}},
  \bibinfo{journal}{{Phys. Fluids}} \textbf{\bibinfo{volume}{{13}}},
  \bibinfo{pages}{2788} (\bibinfo{year}{{2001}}).

\bibitem[{\citenamefont{Usha and Uma}({2002})}]{Usha:ZAngewMathMech02}
\bibinfo{author}{\bibfnamefont{R.}~\bibnamefont{Usha}} \bibnamefont{and}
  \bibinfo{author}{\bibfnamefont{B.}~\bibnamefont{Uma}}, \bibinfo{journal}{{Z.
  Angew. Math. Mech.}} \textbf{\bibinfo{volume}{{82}}}, \bibinfo{pages}{211}
  (\bibinfo{year}{{2002}}).

\bibitem[{\citenamefont{Kitamura et~al.}({2002})\citenamefont{Kitamura,
  Hasegawa, and Yoshizawa}}]{Kitamura:FluidDynRes02}
\bibinfo{author}{\bibfnamefont{A.}~\bibnamefont{Kitamura}},
  \bibinfo{author}{\bibfnamefont{E.}~\bibnamefont{Hasegawa}}, \bibnamefont{and}
  \bibinfo{author}{\bibfnamefont{M.}~\bibnamefont{Yoshizawa}},
  \bibinfo{journal}{{Fluid Dyn. Res.}} \textbf{\bibinfo{volume}{{30}}},
  \bibinfo{pages}{107} (\bibinfo{year}{{2002}}).

\bibitem[{\citenamefont{de~Gennes}(2002)}]{deGennes:EPJE02}
\bibinfo{author}{\bibfnamefont{P.}~\bibnamefont{de~Gennes}},
  \bibinfo{journal}{Eur. Phys. J. E} \textbf{\bibinfo{volume}{7}},
  \bibinfo{pages}{31} (\bibinfo{year}{2002}).

\bibitem[{\citenamefont{Haas and III}(2002)}]{Haas:JMatSci02}
\bibinfo{author}{\bibfnamefont{D.}~\bibnamefont{Haas}} \bibnamefont{and}
  \bibinfo{author}{\bibfnamefont{D.~B.} \bibnamefont{III}},
  \bibinfo{journal}{J. Mat. Sci.} \textbf{\bibinfo{volume}{37}},
  \bibinfo{pages}{2109} (\bibinfo{year}{2002}).

\bibitem[{\citenamefont{Kim et~al.}(2002)\citenamefont{Kim, Yoo, and
  Oh}}]{Kim:JVacSciTechnolB02}
\bibinfo{author}{\bibfnamefont{S.-K.} \bibnamefont{Kim}},
  \bibinfo{author}{\bibfnamefont{J.-Y.} \bibnamefont{Yoo}}, \bibnamefont{and}
  \bibinfo{author}{\bibfnamefont{H.-K.} \bibnamefont{Oh}}, \bibinfo{journal}{J.
  Vac. Sci. Technol. B} \textbf{\bibinfo{volume}{20}}, \bibinfo{pages}{2206}
  (\bibinfo{year}{2002}).

\bibitem[{\citenamefont{Schubert and Dinkel}(2003)}]{Schubert:MatResInnov03}
\bibinfo{author}{\bibfnamefont{D.~W.} \bibnamefont{Schubert}} \bibnamefont{and}
  \bibinfo{author}{\bibfnamefont{T.}~\bibnamefont{Dinkel}},
  \bibinfo{journal}{Mat. Res. Innovat.} \textbf{\bibinfo{volume}{7}},
  \bibinfo{pages}{314} (\bibinfo{year}{2003}).

\bibitem[{\citenamefont{Sisoev et~al.}(2003)\citenamefont{Sisoev, Matar, and
  Lawrence}}]{Sisoev:JChemTechBio03}
\bibinfo{author}{\bibfnamefont{G.}~\bibnamefont{Sisoev}},
  \bibinfo{author}{\bibfnamefont{O.}~\bibnamefont{Matar}}, \bibnamefont{and}
  \bibinfo{author}{\bibfnamefont{C.}~\bibnamefont{Lawrence}},
  \bibinfo{journal}{{J. Chem. Technol. Biotechnol.}}
  \textbf{\bibinfo{volume}{{78}}}, \bibinfo{pages}{151} (\bibinfo{year}{2003}).

\bibitem[{\citenamefont{Sisoev et~al.}({2003})\citenamefont{Sisoev, Matar, and
  Lawrence}}]{Sisoev:JFluidMech03}
\bibinfo{author}{\bibfnamefont{G.}~\bibnamefont{Sisoev}},
  \bibinfo{author}{\bibfnamefont{O.}~\bibnamefont{Matar}}, \bibnamefont{and}
  \bibinfo{author}{\bibfnamefont{C.}~\bibnamefont{Lawrence}},
  \bibinfo{journal}{{J. Fluid Mech.}} \textbf{\bibinfo{volume}{{495}}},
  \bibinfo{pages}{385} (\bibinfo{year}{{2003}}).

\bibitem[{\citenamefont{Dandapat et~al.}({2003})\citenamefont{Dandapat, Daripa,
  and Ray}}]{Dandapat:JApplPhys03}
\bibinfo{author}{\bibfnamefont{B.}~\bibnamefont{Dandapat}},
  \bibinfo{author}{\bibfnamefont{P.}~\bibnamefont{Daripa}}, \bibnamefont{and}
  \bibinfo{author}{\bibfnamefont{P.}~\bibnamefont{Ray}}, \bibinfo{journal}{{J.
  Appl. Phys.}} \textbf{\bibinfo{volume}{{94}}}, \bibinfo{pages}{4144}
  (\bibinfo{year}{{2003}}).

\bibitem[{\citenamefont{Usha and Ravindran}({2004})}]{Usha:IntJNonlinMech04}
\bibinfo{author}{\bibfnamefont{R.}~\bibnamefont{Usha}} \bibnamefont{and}
  \bibinfo{author}{\bibfnamefont{R.}~\bibnamefont{Ravindran}},
  \bibinfo{journal}{{Int. J. Non-Linear Mech.}}
  \textbf{\bibinfo{volume}{{39}}}, \bibinfo{pages}{153}
  (\bibinfo{year}{{2004}}).

\bibitem[{\citenamefont{Schwartz and Roy}({2004})}]{Schwartz:PhysFluids04}
\bibinfo{author}{\bibfnamefont{L.}~\bibnamefont{Schwartz}} \bibnamefont{and}
  \bibinfo{author}{\bibfnamefont{R.}~\bibnamefont{Roy}},
  \bibinfo{journal}{{Phys. Fluids}} \textbf{\bibinfo{volume}{{16}}},
  \bibinfo{pages}{569} (\bibinfo{year}{{2004}}).

\bibitem[{\citenamefont{Matar et~al.}({2004})\citenamefont{Matar, Sisoev, and
  Lawrence}}]{Matar:PhysFluids04}
\bibinfo{author}{\bibfnamefont{O.}~\bibnamefont{Matar}},
  \bibinfo{author}{\bibfnamefont{G.}~\bibnamefont{Sisoev}}, \bibnamefont{and}
  \bibinfo{author}{\bibfnamefont{C.}~\bibnamefont{Lawrence}},
  \bibinfo{journal}{{Phys. Fluids}} \textbf{\bibinfo{volume}{{16}}},
  \bibinfo{pages}{1532} (\bibinfo{year}{{2004}}).

\bibitem[{\citenamefont{Dandapat et~al.}({2005})\citenamefont{Dandapat, Santra,
  and Kitamura}}]{Dandapat:PhysFluids05}
\bibinfo{author}{\bibfnamefont{B.}~\bibnamefont{Dandapat}},
  \bibinfo{author}{\bibfnamefont{B.}~\bibnamefont{Santra}}, \bibnamefont{and}
  \bibinfo{author}{\bibfnamefont{A.}~\bibnamefont{Kitamura}},
  \bibinfo{journal}{{Phys. Fluids}} \textbf{\bibinfo{volume}{{17}}}
  (\bibinfo{year}{{2005}}).

\bibitem[{\citenamefont{Usha et~al.}(2005{\natexlab{a}})\citenamefont{Usha,
  Ravindran, and Uma}}]{Usha:ActaMech05}
\bibinfo{author}{\bibfnamefont{R.}~\bibnamefont{Usha}},
  \bibinfo{author}{\bibfnamefont{R.}~\bibnamefont{Ravindran}},
  \bibnamefont{and} \bibinfo{author}{\bibfnamefont{B.}~\bibnamefont{Uma}},
  \bibinfo{journal}{{Acta Mech.}} \textbf{\bibinfo{volume}{{179}}},
  \bibinfo{pages}{25} (\bibinfo{year}{2005}{\natexlab{a}}).

\bibitem[{\citenamefont{Sisoev et~al.}({2005})\citenamefont{Sisoev, Matar, and
  Lawrence}}]{Sisoev:ChemEngSci05}
\bibinfo{author}{\bibfnamefont{G.}~\bibnamefont{Sisoev}},
  \bibinfo{author}{\bibfnamefont{O.}~\bibnamefont{Matar}}, \bibnamefont{and}
  \bibinfo{author}{\bibfnamefont{C.}~\bibnamefont{Lawrence}},
  \bibinfo{journal}{{Chem. Eng. Sci.}} \textbf{\bibinfo{volume}{{60}}},
  \bibinfo{pages}{2051} (\bibinfo{year}{{2005}}).

\bibitem[{\citenamefont{Momoniat et~al.}({2005})\citenamefont{Momoniat, Myers,
  and Abelman}}]{Momoniat:IntJNonlinMech05}
\bibinfo{author}{\bibfnamefont{E.}~\bibnamefont{Momoniat}},
  \bibinfo{author}{\bibfnamefont{T.}~\bibnamefont{Myers}}, \bibnamefont{and}
  \bibinfo{author}{\bibfnamefont{S.}~\bibnamefont{Abelman}},
  \bibinfo{journal}{{Int. J. Non-Linear Mech.}}
  \textbf{\bibinfo{volume}{{40}}}, \bibinfo{pages}{523}
  (\bibinfo{year}{{2005}}).

\bibitem[{\citenamefont{Wu}({2005})}]{Wu:PhysRevE05}
\bibinfo{author}{\bibfnamefont{L.}~\bibnamefont{Wu}}, \bibinfo{journal}{{Phys.
  Rev. E}} \textbf{\bibinfo{volume}{{72}}} (\bibinfo{year}{{2005}}).

\bibitem[{\citenamefont{Usha et~al.}(2005{\natexlab{b}})\citenamefont{Usha,
  Ravindran, and Uma}}]{Usha:FluidDynRes05}
\bibinfo{author}{\bibfnamefont{R.}~\bibnamefont{Usha}},
  \bibinfo{author}{\bibfnamefont{R.}~\bibnamefont{Ravindran}},
  \bibnamefont{and} \bibinfo{author}{\bibfnamefont{B.}~\bibnamefont{Uma}},
  \bibinfo{journal}{{Fluid Dyn. Res.}} \textbf{\bibinfo{volume}{{37}}},
  \bibinfo{pages}{154} (\bibinfo{year}{2005}{\natexlab{b}}).

\bibitem[{\citenamefont{Usha et~al.}({2005})\citenamefont{Usha, Ravindran, and
  Uma}}]{Usha:PhysFluids05}
\bibinfo{author}{\bibfnamefont{R.}~\bibnamefont{Usha}},
  \bibinfo{author}{\bibfnamefont{R.}~\bibnamefont{Ravindran}},
  \bibnamefont{and} \bibinfo{author}{\bibfnamefont{B.}~\bibnamefont{Uma}},
  \bibinfo{journal}{{Phys. Fluids}} \textbf{\bibinfo{volume}{{17}}}
  (\bibinfo{year}{{2005}}).

\bibitem[{\citenamefont{Myers and Lombe}({2006})}]{Myres:ChemEngProc06}
\bibinfo{author}{\bibfnamefont{T.}~\bibnamefont{Myers}} \bibnamefont{and}
  \bibinfo{author}{\bibfnamefont{M.}~\bibnamefont{Lombe}},
  \bibinfo{journal}{{Chem. Eng. Process.}} \textbf{\bibinfo{volume}{{45}}},
  \bibinfo{pages}{90} (\bibinfo{year}{{2006}}).

\bibitem[{\citenamefont{Wu}({2006})}]{Us:PhysFluids06}
\bibinfo{author}{\bibfnamefont{L.}~\bibnamefont{Wu}}, \bibinfo{journal}{{Phys.
  Fluids}} \textbf{\bibinfo{volume}{{18}}} (\bibinfo{year}{{2006}}).

\bibitem[{\citenamefont{Matar et~al.}({2006})\citenamefont{Matar, Sisoev, and
  Lawrence}}]{Matar:CanJChemEng06}
\bibinfo{author}{\bibfnamefont{O.~K.} \bibnamefont{Matar}},
  \bibinfo{author}{\bibfnamefont{G.~M.} \bibnamefont{Sisoev}},
  \bibnamefont{and} \bibinfo{author}{\bibfnamefont{C.~J.}
  \bibnamefont{Lawrence}}, \bibinfo{journal}{{Can. J. Chem. Eng.}}
  \textbf{\bibinfo{volume}{{84}}}, \bibinfo{pages}{625}
  (\bibinfo{year}{{2006}}).

\bibitem[{\citenamefont{Tabuteau et~al.}({2007})\citenamefont{Tabuteau, Baudez,
  Chateau, and Coussot}}]{Tabuteau:RheolActa07}
\bibinfo{author}{\bibfnamefont{H.}~\bibnamefont{Tabuteau}},
  \bibinfo{author}{\bibfnamefont{J.~C.} \bibnamefont{Baudez}},
  \bibinfo{author}{\bibfnamefont{X.}~\bibnamefont{Chateau}}, \bibnamefont{and}
  \bibinfo{author}{\bibfnamefont{P.}~\bibnamefont{Coussot}},
  \bibinfo{journal}{{Rheol. Acta}} \textbf{\bibinfo{volume}{{46}}},
  \bibinfo{pages}{341} (\bibinfo{year}{{2007}}).

\bibitem[{\citenamefont{Charpin et~al.}(2007)\citenamefont{Charpin, Lombe, and
  Myers}}]{Charpin:PhysRevE07}
\bibinfo{author}{\bibfnamefont{J.~P.~F.} \bibnamefont{Charpin}},
  \bibinfo{author}{\bibfnamefont{M.}~\bibnamefont{Lombe}}, \bibnamefont{and}
  \bibinfo{author}{\bibfnamefont{T.~G.} \bibnamefont{Myers}},
  \bibinfo{journal}{Phys. Rev. E} \textbf{\bibinfo{volume}{76}},
  \bibinfo{pages}{016312} (\bibinfo{year}{2007}).

\bibitem[{\citenamefont{Wu}({2007})}]{Wu:SensActuatorA07}
\bibinfo{author}{\bibfnamefont{L.}~\bibnamefont{Wu}}, \bibinfo{journal}{{Sens.
  Actuator A-Phys.}} \textbf{\bibinfo{volume}{{134}}}, \bibinfo{pages}{140}
  (\bibinfo{year}{{2007}}).

\bibitem[{\citenamefont{Holloway et~al.}({2007})\citenamefont{Holloway, Habdas,
  Semsarillar, Burfitt, and de~Bruyn}}]{Holloway:PhysRevE07}
\bibinfo{author}{\bibfnamefont{K.~E.} \bibnamefont{Holloway}},
  \bibinfo{author}{\bibfnamefont{P.}~\bibnamefont{Habdas}},
  \bibinfo{author}{\bibfnamefont{N.}~\bibnamefont{Semsarillar}},
  \bibinfo{author}{\bibfnamefont{K.}~\bibnamefont{Burfitt}}, \bibnamefont{and}
  \bibinfo{author}{\bibfnamefont{J.~R.} \bibnamefont{de~Bruyn}},
  \bibinfo{journal}{{Phys. Rev. E}} \textbf{\bibinfo{volume}{{75}}}
  (\bibinfo{year}{{2007}}).

\bibitem[{\citenamefont{Cregan and O'Brien}(2007)}]{Cregan:JCollInterfSci07}
\bibinfo{author}{\bibfnamefont{V.}~\bibnamefont{Cregan}} \bibnamefont{and}
  \bibinfo{author}{\bibfnamefont{S.}~\bibnamefont{O'Brien}},
  \bibinfo{journal}{J. Colloid Interface Sci.} \textbf{\bibinfo{volume}{314}},
  \bibinfo{pages}{324} (\bibinfo{year}{2007}).

\bibitem[{\citenamefont{Zhao and Marshall}({2008})}]{Zhao:PhysFluids08}
\bibinfo{author}{\bibfnamefont{Y.}~\bibnamefont{Zhao}} \bibnamefont{and}
  \bibinfo{author}{\bibfnamefont{J.~S.} \bibnamefont{Marshall}},
  \bibinfo{journal}{{Phys. Fluids}} \textbf{\bibinfo{volume}{{20}}}
  (\bibinfo{year}{{2008}}).

\bibitem[{\citenamefont{Matar et~al.}({2008})\citenamefont{Matar, Sisoev, and
  Lawrence}}]{Matar:ChemEngSci08}
\bibinfo{author}{\bibfnamefont{O.~K.} \bibnamefont{Matar}},
  \bibinfo{author}{\bibfnamefont{G.~M.} \bibnamefont{Sisoev}},
  \bibnamefont{and} \bibinfo{author}{\bibfnamefont{C.~J.}
  \bibnamefont{Lawrence}}, \bibinfo{journal}{{Chem. Eng. Sci.}}
  \textbf{\bibinfo{volume}{{63}}}, \bibinfo{pages}{2225}
  (\bibinfo{year}{{2008}}).

\bibitem[{\citenamefont{Chen and Lin}(2009)}]{Chen:MathProbEng09}
\bibinfo{author}{\bibfnamefont{C.-K.} \bibnamefont{Chen}} \bibnamefont{and}
  \bibinfo{author}{\bibfnamefont{M.-C.} \bibnamefont{Lin}},
  \bibinfo{journal}{Math. Prob. Eng.} \textbf{\bibinfo{volume}{2009}},
  \bibinfo{pages}{948672} (\bibinfo{year}{2009}).

\bibitem[{\citenamefont{Mukhopadhyay and
  Behringer}({2009})}]{Mukhopadhyay:JPhysCondMat09}
\bibinfo{author}{\bibfnamefont{S.}~\bibnamefont{Mukhopadhyay}}
  \bibnamefont{and} \bibinfo{author}{\bibfnamefont{R.~P.}
  \bibnamefont{Behringer}}, \bibinfo{journal}{{J. Phys.-Condes. Matter}}
  \textbf{\bibinfo{volume}{{21}}} (\bibinfo{year}{{2009}}).

\bibitem[{\citenamefont{Chen and Lai}({2010})}]{Chen:MathProblEng10:2}
\bibinfo{author}{\bibfnamefont{C.-K.} \bibnamefont{Chen}} \bibnamefont{and}
  \bibinfo{author}{\bibfnamefont{D.-Y.} \bibnamefont{Lai}},
  \bibinfo{journal}{{Math. Probl. Eng.}}  (\bibinfo{year}{{2010}}).

\bibitem[{\citenamefont{Chen and Lai}(2010)}]{Chen:MathProblEng10:1}
\bibinfo{author}{\bibfnamefont{C.-K.} \bibnamefont{Chen}} \bibnamefont{and}
  \bibinfo{author}{\bibfnamefont{D.-Y.} \bibnamefont{Lai}},
  \bibinfo{journal}{Math. Prob. Eng.} \textbf{\bibinfo{volume}{2010}},
  \bibinfo{pages}{987981} (\bibinfo{year}{2010}).

\bibitem[{\citenamefont{Jung et~al.}(2010)\citenamefont{Jung, Kang, and
  Koo}}]{Jung:IntJHMT10}
\bibinfo{author}{\bibfnamefont{J.-Y.} \bibnamefont{Jung}},
  \bibinfo{author}{\bibfnamefont{Y.}~\bibnamefont{Kang}}, \bibnamefont{and}
  \bibinfo{author}{\bibfnamefont{J.}~\bibnamefont{Koo}}, \bibinfo{journal}{Int.
  J. Heat Mass Transf.} \textbf{\bibinfo{volume}{53}}, \bibinfo{pages}{1712}
  (\bibinfo{year}{2010}).

\bibitem[{\citenamefont{McIntyre and Brush}(2010)}]{McIntyre:JFluidMech10}
\bibinfo{author}{\bibfnamefont{A.}~\bibnamefont{McIntyre}} \bibnamefont{and}
  \bibinfo{author}{\bibfnamefont{L.}~\bibnamefont{Brush}}, \bibinfo{journal}{J.
  Fluid Mech.} \textbf{\bibinfo{volume}{647}}, \bibinfo{pages}{265}
  (\bibinfo{year}{2010}).

\bibitem[{\citenamefont{Temple-Boyer et~al.}(2010)\citenamefont{Temple-Boyer,
  Mazenq, Doucet, Conedera, Torbiero, and Launay}}]{Temple-Boyer:MicroEng10}
\bibinfo{author}{\bibfnamefont{P.}~\bibnamefont{Temple-Boyer}},
  \bibinfo{author}{\bibfnamefont{L.}~\bibnamefont{Mazenq}},
  \bibinfo{author}{\bibfnamefont{J.}~\bibnamefont{Doucet}},
  \bibinfo{author}{\bibfnamefont{V.}~\bibnamefont{Conedera}},
  \bibinfo{author}{\bibfnamefont{B.}~\bibnamefont{Torbiero}}, \bibnamefont{and}
  \bibinfo{author}{\bibfnamefont{J.}~\bibnamefont{Launay}},
  \bibinfo{journal}{Microelectr. Eng.} \textbf{\bibinfo{volume}{87}},
  \bibinfo{pages}{163} (\bibinfo{year}{2010}).

\bibitem[{\citenamefont{Muench et~al.}(2011)\citenamefont{Muench, Please, and
  Wagner}}]{Muench:PhysFluids11}
\bibinfo{author}{\bibfnamefont{A.}~\bibnamefont{Muench}},
  \bibinfo{author}{\bibfnamefont{C.~P.} \bibnamefont{Please}},
  \bibnamefont{and} \bibinfo{author}{\bibfnamefont{B.}~\bibnamefont{Wagner}},
  \bibinfo{journal}{Phys. Fluids} \textbf{\bibinfo{volume}{23}},
  \bibinfo{pages}{102101} (\bibinfo{year}{2011}).

\bibitem[{\citenamefont{Modhien and Momoniat}({2011})}]{Modhien:ApplMathMod11}
\bibinfo{author}{\bibfnamefont{N.}~\bibnamefont{Modhien}} \bibnamefont{and}
  \bibinfo{author}{\bibfnamefont{E.}~\bibnamefont{Momoniat}},
  \bibinfo{journal}{{Appl. Math. Model.}} \textbf{\bibinfo{volume}{{35}}},
  \bibinfo{pages}{1264} (\bibinfo{year}{{2011}}).

\bibitem[{\citenamefont{Chen et~al.}({2011})\citenamefont{Chen, Lin, and
  Chen}}]{Chen:JMech11}
\bibinfo{author}{\bibfnamefont{C.~K.} \bibnamefont{Chen}},
  \bibinfo{author}{\bibfnamefont{M.~C.} \bibnamefont{Lin}}, \bibnamefont{and}
  \bibinfo{author}{\bibfnamefont{C.~I.} \bibnamefont{Chen}},
  \bibinfo{journal}{{J. Mech.}} \textbf{\bibinfo{volume}{{27}}},
  \bibinfo{pages}{95} (\bibinfo{year}{{2011}}).

\bibitem[{\citenamefont{Dandapat and
  Singh}({2011})}]{Dandapat:IntJNonlinMech11}
\bibinfo{author}{\bibfnamefont{B.~S.} \bibnamefont{Dandapat}} \bibnamefont{and}
  \bibinfo{author}{\bibfnamefont{S.~K.} \bibnamefont{Singh}},
  \bibinfo{journal}{{Int. J. Non-Linear Mech.}}
  \textbf{\bibinfo{volume}{{46}}}, \bibinfo{pages}{272}
  (\bibinfo{year}{{2011}}).

\bibitem[{\citenamefont{Dandapat and Singh}({2012})}]{Dandapat:CommNonlinSci12}
\bibinfo{author}{\bibfnamefont{B.~S.} \bibnamefont{Dandapat}} \bibnamefont{and}
  \bibinfo{author}{\bibfnamefont{S.~K.} \bibnamefont{Singh}},
  \bibinfo{journal}{{Commun. Nonlinear Sci. Numer. Simul.}}
  \textbf{\bibinfo{volume}{{17}}}, \bibinfo{pages}{2854}
  (\bibinfo{year}{{2012}}).

\bibitem[{\citenamefont{Lin and Chen}({2012})}]{Lin:ApplMathMod12}
\bibinfo{author}{\bibfnamefont{M.~C.} \bibnamefont{Lin}} \bibnamefont{and}
  \bibinfo{author}{\bibfnamefont{C.~K.} \bibnamefont{Chen}},
  \bibinfo{journal}{{Appl. Math. Model.}} \textbf{\bibinfo{volume}{{36}}},
  \bibinfo{pages}{2536} (\bibinfo{year}{{2012}}).

\bibitem[{\citenamefont{Norrman et~al.}(2005)\citenamefont{Norrman,
  Ghanbari-Siahkali, and Larsen}}]{Norrman:AnRepProcChemC05}
\bibinfo{author}{\bibfnamefont{K.}~\bibnamefont{Norrman}},
  \bibinfo{author}{\bibfnamefont{A.}~\bibnamefont{Ghanbari-Siahkali}},
  \bibnamefont{and} \bibinfo{author}{\bibfnamefont{N.~B.}
  \bibnamefont{Larsen}}, \bibinfo{journal}{Annu. Rep. Prog. Chem. C}
  \textbf{\bibinfo{volume}{101}}, \bibinfo{pages}{174} (\bibinfo{year}{2005}),
  \bibinfo{note}{in particular references 25 through 37 cited therein}.

\bibitem[{\citenamefont{Riegler and K\"ohler}(2007)}]{Riegler:NatPhys07}
\bibinfo{author}{\bibfnamefont{H.}~\bibnamefont{Riegler}} \bibnamefont{and}
  \bibinfo{author}{\bibfnamefont{R.}~\bibnamefont{K\"ohler}},
  \bibinfo{journal}{Nat. Phys.} \textbf{\bibinfo{volume}{3}},
  \bibinfo{pages}{890} (\bibinfo{year}{2007}).

\bibitem[{\citenamefont{Berg et~al.}(2010)\citenamefont{Berg, Weber, and
  Riegler}}]{Berg:PhysRevLett10}
\bibinfo{author}{\bibfnamefont{J.~K.} \bibnamefont{Berg}},
  \bibinfo{author}{\bibfnamefont{C.~M.} \bibnamefont{Weber}}, \bibnamefont{and}
  \bibinfo{author}{\bibfnamefont{H.}~\bibnamefont{Riegler}},
  \bibinfo{journal}{Phys. Rev. Lett.} \textbf{\bibinfo{volume}{105}},
  \bibinfo{pages}{076103} (\bibinfo{year}{2010}).

\bibitem[{\citenamefont{Brookshier et~al.}(1999)\citenamefont{Brookshier,
  Chusuei, and Goodman}}]{Brookshier:Langmuir99}
\bibinfo{author}{\bibfnamefont{M.~A.} \bibnamefont{Brookshier}},
  \bibinfo{author}{\bibfnamefont{C.~C.} \bibnamefont{Chusuei}},
  \bibnamefont{and} \bibinfo{author}{\bibfnamefont{D.~W.}
  \bibnamefont{Goodman}}, \bibinfo{journal}{Langmuir}
  \textbf{\bibinfo{volume}{15}}, \bibinfo{pages}{2043} (\bibinfo{year}{1999}).

\bibitem[{\citenamefont{Rabani et~al.}(2003)\citenamefont{Rabani, Reichman,
  Geissler, and Brus}}]{Rabani:Nature03}
\bibinfo{author}{\bibfnamefont{E.}~\bibnamefont{Rabani}},
  \bibinfo{author}{\bibfnamefont{D.~R.} \bibnamefont{Reichman}},
  \bibinfo{author}{\bibfnamefont{P.~L.} \bibnamefont{Geissler}},
  \bibnamefont{and} \bibinfo{author}{\bibfnamefont{L.~E.} \bibnamefont{Brus}},
  \bibinfo{journal}{Nature} \textbf{\bibinfo{volume}{426}},
  \bibinfo{pages}{271} (\bibinfo{year}{2003}).

\bibitem[{\citenamefont{Bigioni et~al.}(2006)\citenamefont{Bigioni, Lin,
  Nguyen, Corwin, Witten, and Jaeger}}]{Bigioni:NatMat06}
\bibinfo{author}{\bibfnamefont{T.~P.} \bibnamefont{Bigioni}},
  \bibinfo{author}{\bibfnamefont{X.-M.} \bibnamefont{Lin}},
  \bibinfo{author}{\bibfnamefont{T.~T.} \bibnamefont{Nguyen}},
  \bibinfo{author}{\bibfnamefont{E.~I.} \bibnamefont{Corwin}},
  \bibinfo{author}{\bibfnamefont{T.~A.} \bibnamefont{Witten}},
  \bibnamefont{and} \bibinfo{author}{\bibfnamefont{H.~M.}
  \bibnamefont{Jaeger}}, \bibinfo{journal}{Nat. Mater.}
  \textbf{\bibinfo{volume}{5}}, \bibinfo{pages}{265} (\bibinfo{year}{2006}).

\bibitem[{\citenamefont{Hanrath et~al.}(2009)\citenamefont{Hanrath, Choi, and
  Smilgies}}]{Hanrath:Nano09}
\bibinfo{author}{\bibfnamefont{T.}~\bibnamefont{Hanrath}},
  \bibinfo{author}{\bibfnamefont{J.~J.} \bibnamefont{Choi}}, \bibnamefont{and}
  \bibinfo{author}{\bibfnamefont{D.-M.} \bibnamefont{Smilgies}},
  \bibinfo{journal}{ACS Nano} \textbf{\bibinfo{volume}{3}},
  \bibinfo{pages}{2975} (\bibinfo{year}{2009}).

\bibitem[{\citenamefont{Heitsch et~al.}(2010)\citenamefont{Heitsch, Patel,
  Goodfellow, Smilgies, and Korgel}}]{Heitsch:JPhysChem10}
\bibinfo{author}{\bibfnamefont{A.~T.} \bibnamefont{Heitsch}},
  \bibinfo{author}{\bibfnamefont{R.~N.} \bibnamefont{Patel}},
  \bibinfo{author}{\bibfnamefont{B.~W.} \bibnamefont{Goodfellow}},
  \bibinfo{author}{\bibfnamefont{D.-M.} \bibnamefont{Smilgies}},
  \bibnamefont{and} \bibinfo{author}{\bibfnamefont{B.}~\bibnamefont{Korgel}},
  \bibinfo{journal}{J. Phys. Chem. C} \textbf{\bibinfo{volume}{114}},
  \bibinfo{pages}{1616} (\bibinfo{year}{2010}).

\bibitem[{\citenamefont{Klecha et~al.}(2010)\citenamefont{Klecha, Ingert, and
  Pileni}}]{Klecha:JPhysChemLett10}
\bibinfo{author}{\bibfnamefont{E.}~\bibnamefont{Klecha}},
  \bibinfo{author}{\bibfnamefont{D.}~\bibnamefont{Ingert}}, \bibnamefont{and}
  \bibinfo{author}{\bibfnamefont{M.~P.} \bibnamefont{Pileni}},
  \bibinfo{journal}{J. Phys. Chem. Lett.} \textbf{\bibinfo{volume}{1}},
  \bibinfo{pages}{14427} (\bibinfo{year}{2010}).

\bibitem[{\citenamefont{Johnston-Peck et~al.}(2011)\citenamefont{Johnston-Peck,
  Wang, and Tracy}}]{Johnston-Peck:Langmuir11}
\bibinfo{author}{\bibfnamefont{A.~C.} \bibnamefont{Johnston-Peck}},
  \bibinfo{author}{\bibfnamefont{J.}~\bibnamefont{Wang}}, \bibnamefont{and}
  \bibinfo{author}{\bibfnamefont{J.~B.} \bibnamefont{Tracy}},
  \bibinfo{journal}{Langmuir} \textbf{\bibinfo{volume}{27}},
  \bibinfo{pages}{5040} (\bibinfo{year}{2011}).

\bibitem[{\citenamefont{Marin et~al.}(2011)\citenamefont{Marin, Gelderblom,
  Lohse, and Snoeijer}}]{Marin:PhysRevLett11}
\bibinfo{author}{\bibfnamefont{A.~G.} \bibnamefont{Marin}},
  \bibinfo{author}{\bibfnamefont{H.}~\bibnamefont{Gelderblom}},
  \bibinfo{author}{\bibfnamefont{D.}~\bibnamefont{Lohse}}, \bibnamefont{and}
  \bibinfo{author}{\bibfnamefont{J.~H.} \bibnamefont{Snoeijer}},
  \bibinfo{journal}{Phys. Rev. Lett.} \textbf{\bibinfo{volume}{107}},
  \bibinfo{pages}{085502} (\bibinfo{year}{2011}).

\bibitem[{\citenamefont{Rauch and Kohler}(2002)}]{Rauch:PhysRevLett02}
\bibinfo{author}{\bibfnamefont{J.}~\bibnamefont{Rauch}} \bibnamefont{and}
  \bibinfo{author}{\bibfnamefont{W.}~\bibnamefont{Kohler}},
  \bibinfo{journal}{Phys. Rev. Lett.} \textbf{\bibinfo{volume}{88}},
  \bibinfo{pages}{185901} (\bibinfo{year}{2002}).

\end{thebibliography}
\end{document}